\documentclass[physrev, twocolumn, nobibnotes, superscriptaddress, notitlepage, noeprint,prapplied]{revtex4-2}

\usepackage{float}
\usepackage{graphicx}
\usepackage{mathrsfs}
\usepackage{overpic}
\usepackage{wrapfig}
\usepackage{braket}
\usepackage{color}
\usepackage{verbatim}
\usepackage{amssymb,amsmath,mathtools}
\usepackage{upgreek}
\usepackage{bbold}
\usepackage{cancel}
\usepackage{textgreek}
\usepackage[normalem]{ulem}
\newcommand{\specialcell}[2][c]{%
  \begin{tabular}[#1]{@{}c@{}}#2\end{tabular}}
\usepackage[usenames,dvipsnames]{xcolor}
\usepackage{hyperref}
\hypersetup{linktoc=page,
colorlinks,
urlcolor=RedViolet,
citecolor=RedViolet,
linkcolor=RedViolet}

\begin{document}
\preprint{APS/123-QED}
\title{Prospects for Ultralow-Mass Nuclear Magnetic Resonance using Spin Defects in Hexagonal Boron Nitride}

\author{Declan M. Daly}
\altaffiliation{Equal contribution}
\affiliation{Department of Physics, University of Maryland, College Park, MD, USA}

\author{Niko R. Reed}
\altaffiliation{Equal contribution}
\affiliation{Department of Physics, University of Maryland, College Park, MD, USA}
\affiliation{Quantum Technology Center, University of Maryland, College Park, MD, USA}
\affiliation{Joint Quantum Institute, University of Maryland, College Park, MD, USA}

\author{Stephen J. DeVience}
\affiliation{Quantum Technology Center, University of Maryland, College Park, MD, USA}

\author{Zechuan Yin}
\affiliation{Quantum Technology Center, University of Maryland, College Park, MD, USA}
\affiliation{Department of Electrical and Computer Engineering, University of Maryland, College Park, MD, USA}

\author{Johannes Cremer}
\affiliation{Quantum Technology Center, University of Maryland, College Park, MD, USA}

\author{Andrew J. Beling}
\affiliation{Department of Physics, University of Maryland, College Park, MD, USA}
\affiliation{Quantum Technology Center, University of Maryland, College Park, MD, USA}

\author{John W. Blanchard}
\affiliation{Quantum Technology Center, University of Maryland, College Park, MD, USA}

\author{Ronald L. Walsworth}
\email{walsworth@umd.edu}
\affiliation{Department of Physics, University of Maryland, College Park, MD, USA}
\affiliation{Quantum Technology Center, University of Maryland, College Park, MD, USA}
\affiliation{Joint Quantum Institute, University of Maryland, College Park, MD, USA}
\affiliation{Department of Electrical and Computer Engineering, University of Maryland, College Park, MD, USA}

\date{May 1st, 2025}

\begin{abstract}
Optically active quantum defects in solids, such as the nitrogen vacancy (NV) center in diamond, are a leading modality for micron-scale and nanoscale (ultralow-mass) nuclear magnetic resonance (NMR) spectroscopy and imaging under ambient conditions. 
However, the spin and optical properties of NV centers degrade when closer than about 10 nm from the diamond surface, limiting NMR sensitivity as well as spectral and spatial resolution. 
Here we outline efforts to develop an alternative nanoscale NMR sensor using the negatively charged boron vacancy ($V_B^-$) in hexagonal boron nitride (hBN). 
As a van der Waals material, hBN's surface is free from dangling bonds and other sources of paramagnetic noise that degrade the performance of near surface NVs, allowing stable $V_B^-$ defects to exist $\sim1\,$nm from the material surface. 
We discuss the properties of boron vacancies as they apply to narrowband (AC) magnetic field sensing and outline experimental designs optimized for this system. 
We propose measurement protocols for $V_B^-$ NMR for both statistically and uniformly polarized samples at the nano- and micron-scales, including relevant pulse sequences, sensitivity calculations, and sample confinement strategies; and compare the expected performance to NV-NMR. 
We estimate back-action effects between the $V_B^-$ electronic spins and the sample nuclear spins at the nanoscale; and account for unconventional diffusion dynamics in the flow-restricted nanoscale regime, calculating its effects on the expected $V_B^-$ NMR signal. 
Lastly, we identify potential sample targets and operational regimes best suited for both nanoscale and micron-scale $V_B^-$ NMR.
\end{abstract}

\maketitle

\section{Introduction}

Nuclear Magnetic Resonance (NMR) spectroscopy is a leading technique for chemical identification and structural analysis, with wide-ranging applications in the physical and life sciences \cite{theillet_-cell_2022, wishart_nmr_2022, jonas_prediction_2022}. Conventional NMR signal detection uses inductive readout, e.g., measuring the free induction decay (FID) of nuclear spins in a sample \cite{abraham_introduction_1988}.  However, the signal-to-noise ratio (SNR) of inductive readout scales as the ratio of sample volume to Johnson noise in the coil \cite{ali_design_2017}, such that NMR signals are typically undetectable for small volumes ($\lesssim$ nanoliter) and for thin surfaces at room temperature \cite{herzog_boundary_2014, grisi_nmr_2017}.  Although nuclear spin hyperpolarization strategies are capable of providing SNR boosts of several orders of magnitude to enable surface-selective NMR spectroscopy \cite{DNPSENS}, these are limited to relatively large ($>10$\,mg) samples of mesoporous or nanoparticulate materials having surface areas of at least 10s of m$^2/$g.
By crushing flat surfaces into a fine powder, it is possible to reach surface areas down to $0.01\,$m$^2$/g with optimized experimental parameters \cite{Walder2019}, but measurement of a single surface remains far beyond reach, necessitating the development of new NMR techniques for such applications. Conventional detection limitations also restrict the use of NMR for ultralow-mass applications where large volumes of samples are difficult to obtain such as pharmaceutical development, measurement of biological metabolites from single cells, and the study of irreplaceable samples \cite{molinskiNMRNaturalProducts2010,woltersMicroscaleNMR2002}. 

This size limitation can be overcome by replacing inductive NMR signal detection with magnetically sensitive, optically-probed quantum defects in solids, e.g., nitrogen vacancy (NV) centers in diamond \cite{mamin_nanoscale_2013, staudacher_nuclear_2013,lovchinsky_nuclear_2016}.  While NV centers have demonstrated impressive results in both nanoscale and micron-scale NMR, \cite{glenn_high-resolution_2018, pham_nmr_2016, aslam_nanoscale_2017, bruckmaier_imaging_2023, bucher_hyperpolarization-enhanced_2020, schwartz_blueprint_2019}, their potential sensitivity is limited by standoff distance from the sample. In particular, NV centers must be created at a minimum depth of several nanometers from the diamond surface to maintain desirable coherence properties, as surface effects degrade defect stability. As the strength of dipolar coupling between the sample and sensor spins falls off as $r^{-3}$, even a few nanometers can critically impact a nanoscale NMR experiment equivalent to several orders of magnitude in SNR. 

\begin{figure}
    
    \includegraphics[width=1\linewidth]{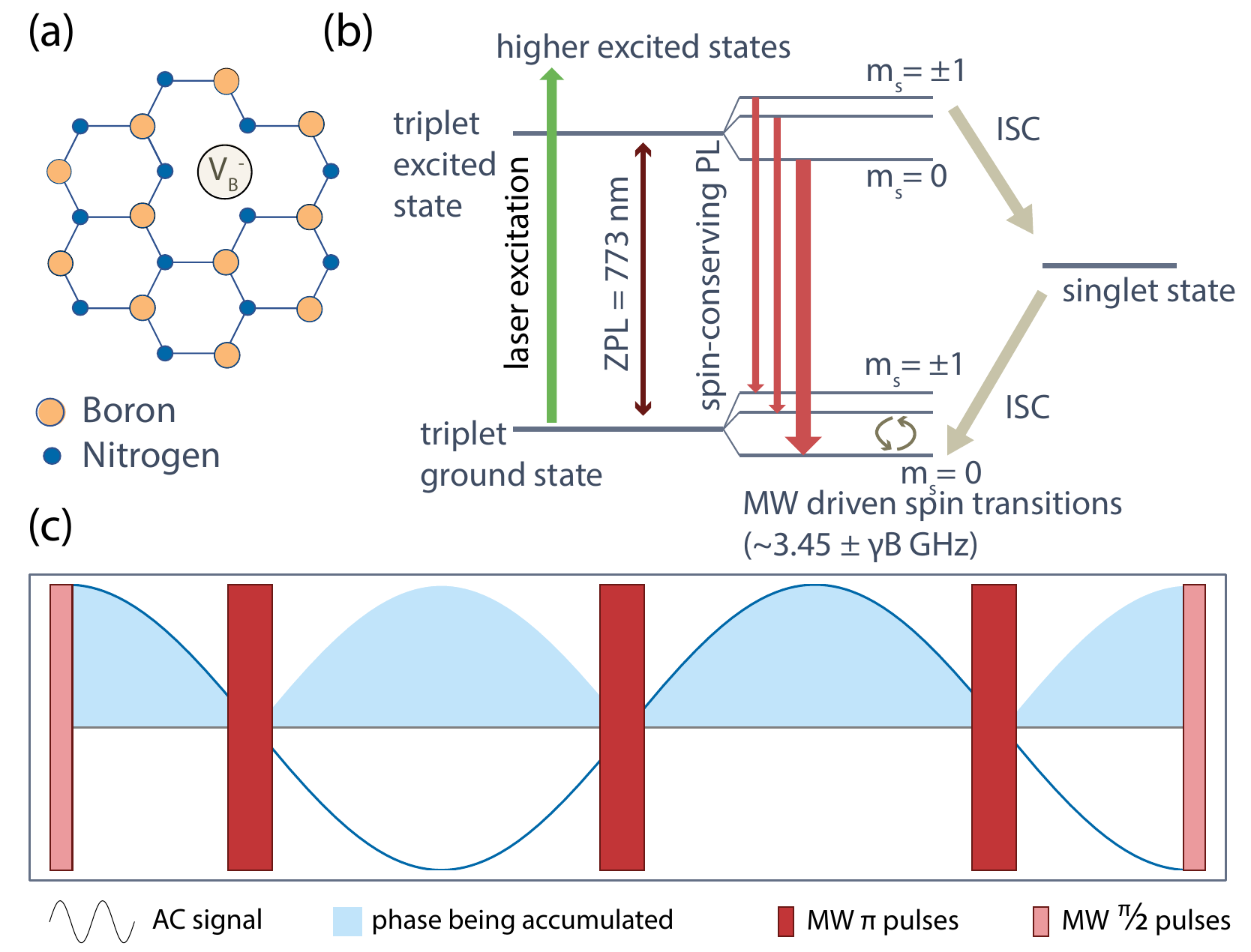}
    \caption{\label{introfig} (a) Atomic structure of the boron vacancy defect ($V_B^-$) within the hBN lattice. (b) $V_B^-$ energy level structure. A green or blue laser excites electronic spins (S=1) from the triplet ground state to the triplet excited state. Spins in the $m_s=0$ ground state that are brought to the excited state radiatively decay back to the $m_s=0$ ground state, emitting infrared photoluminescence (PL).  The difference in absorption and emission wavelengths is governed by spin-phonon interactions.  A portion of spins excited from the $m_s=\pm 1$ ground state will exhibit nonradiative decay to $m_s=0$ through the intersystem crossing. In the absence of microwaves (MW), this process with spin-selective PL simultaneously prepares an ensemble of spins in the $m_s=0$ state and allows for spin-state dependent readout; with a large number of defects, the average projection onto the $m_s=0$ state is encoded in the strength of infrared PL. Note that the energy level structure and response to optical pumping of $V_B^-$ is similar to that of NV centers, allowing similar experimental protocols to be used. (c) Schematic of phase accumulation by defect spins in the presence of an AC signal and a measurement protocol involving periodic MW $\pi$-pulses resonant with the sensor spin.}

\end{figure}

Negatively charged boron vacancies ($V^{-}_B$) in hexagonal boron nitride (hBN), a van der Waals material, are promising candidates for ultralow-mass NMR due to their magnetic sensitivity and stability near the hBN surface \cite{rizzato_extending_2023,vaidya_quantum_2023,durand_optically_2023}. Like NV centers, $V^{-}_B$ are electronic spin-1 defects that can be optically initialized and read out at room temperature (see Fig. \ref{introfig}). \cite{gottscholl_room_2021}. Like diamond, hBN is chemically and thermally inert, making it suitable for a wide range of experimental conditions \cite{kim_color_2023}. hBN has been used as an encapsulating layer in many systems of interest for NMR \cite{seitz_long-term_2019,huang_encapsulation_2022}, including the fabrication of nanowells to enclose very small sample volumes, which offers utility for ultralow-mass NMR \cite{clark_tracking_2022,keerthi_water_2021,pagliero_slow_2024}. As $V^{-}_B$ are stable within a single atomic layer from the hBN surface, sensor-sample standoff distances as small as 1\,nm may be possible \cite{durand_optically_2023}.

\begin{figure*}
    
    \includegraphics[width=1\linewidth]{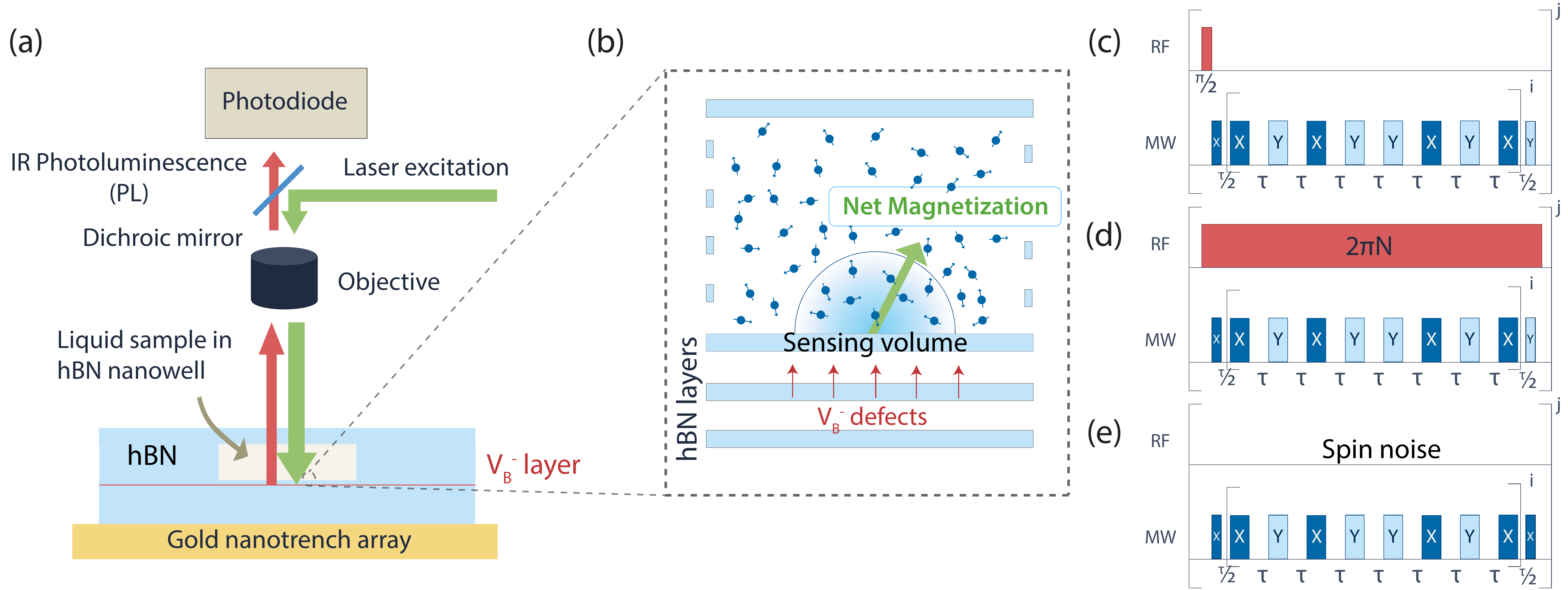}
    \caption{\label{introfig2b} (a) Proposed experimental design for nanoscale $V_B^-$ NMR, inspired by NV-NMR systems, able to leverage special properties of hBN. An objective is used for both $V_B^-$ illumination and PL collection.  For liquid samples, the experimental target is contained within encapsulating layers of hBN, confined within a nanowell to limit effects of diffusion.  A gold nanotrench array (NTA) below the hBN layer enhances MW Rabi driving of the defect spins, allowing for high-frequency NMR measurements to accumulate relevant signal phase. (b) Cartoon showing the net magnetization of a statistically polarized sample in the hBN nanowell.  The effective sample volume of a single $V_B^-$ defect may contain only a handful of nuclear spins.  In such a small population, the net polarization is typically much greater than that of a thermally polarized sample due to the stochastic nature of the spin distribution (i.e. due to statistical polarization fluctuations).  (c) Generalized pulse protocol for defect NMR with transverse magnetization where radiofrequency (RF) driving controls sample nuclear spins and MW driving controls sensor electronic spins. In this protocol, \textit{i} refers to the number of XY8 repetitions within a single acquisition, such as XY8-2 or XY8-3.  \textit{j} refers to the number of acquisitions, where \textit{j} depends on whether XY8-k protocols, correlation spectroscopy, or CASR is used. (d) Pulse protocol for defect NMR with longitudinal detection. Here, the NMR signal $\langle M_y \rangle$  is encoded in the sample's $\langle M_z \rangle$ amplitude for each acquisition, e.g., using the AERIS \cite{munuera-javaloy_high-resolution_2023} or DRACAERIS \cite{daly_nutation-based_2024} protocols. This approach allows the NMR signal to be downconverted to a frequency measurable by the defect sensor ($\sim$ few MHz), which is most relevant for high-field NMR applications. (e) Pulse protocol for defect NMR with statistical sample magnetization. For these protocols RF control is not needed, as nuclear spin noise is directly detected.}

\end{figure*}

In this paper, we outline a strategy for $V^{-}_B$ NMR experiments (see Fig. \ref{introfig2b}) and assess the expected performance. We review parameters relevant to narrowband (AC) magnetic sensing, suggest experimental designs, and perform an AC sensitivity comparison between $V^{-}_B$ and NV centers at both the nanoscale and micron-scale (see Table \ref{table1}). We identify key application spaces for $V^{-}_B$ NMR, including sample targets for potential high-impact investigations, and highlight regimes where ensemble $V_B^-$ NMR approaches could outperform current NV-NMR results.

\begin{table*}
\begin{ruledtabular}
\begin{tabular}{|c|cccc|}
& \specialcell{\textbf{Number of Spins}} & \specialcell{\textbf{Polarization}} & \specialcell{\textbf{Recommended Protocols}} & \specialcell{\textbf{Characteristics}} \\
\hline
\specialcell{\textbf{Strong Sensor/} \\ \textbf{Sample Spin} \\ \textbf{Coupling}}& \specialcell{$\sim1$\\ (nm scale)}& \specialcell{Not required; spins \\ are detected individually} & \specialcell{XY8 \\ Correlation Spectroscopy} & \specialcell{Extraction of physical, \\ chemical, and spatial \\ information about \\ individual molecules}\\
\hline
\specialcell{\textbf{Statistical} \\ \textbf{Polarization}} & \specialcell{$<10^6$\\(nm scale)} & Statistical fluctuations & \specialcell{XY8 \\ Correlation Spectroscopy} & \specialcell{Extremely mass limited \\samples; 2D materials, \\ liquid crystals. Emphasis \\ on studying crystal \\ structure and deformation, \\order parameters} \\
\hline
\specialcell{\textbf{Bulk Defect} \\ \textbf{NMR}} & \specialcell{$10^6-10^{13}$ \\ ($\mu$m scale)} & \specialcell{Thermal, \\ Hyperpolarization} & \specialcell{CASR, DRACAERIS} & \specialcell{Study of mass limited \\ samples due to expense or \\ scarcity, e.g., pharmaceutical \\ development. Emphasis on \\ chemical identification} \\
\hline
\specialcell{\textbf{Conventional} \\ \textbf{NMR}} & \specialcell{$>10^{13}$ \\ (mm and above)}& \specialcell{Thermal, \\ Hyperpolarization} & Conventional NMR & \specialcell{Precision measurements \\of bulk samples} \\

\end{tabular}
\end{ruledtabular}
\caption{\label{table1} Overview of NMR regimes, sorted by approximate number of sample spins. Boron vacancies ($V_B^-$) are best suited to interrogate statistically polarized samples, while remaining competitive with NVs in the regimes of strong sensor/sample spin coupling or bulk defect NMR.}
\end{table*}

\section{VB- Sensitivity for NMR} \label{sensitivity}

In order to probe NMR regimes inaccessible to inductive techniques, a quantum sensor must be able to detect nanotesla or even picotesla AC magnetic signals. Thus, we first evaluate the AC sensitivity of both $V_B^-$ and NV defects in the context of existing NV-NMR protocols.

While AC sensing with $V_B^-$ has only recently been demonstrated \cite{rizzato_extending_2023}, the underlying physical mechanisms operate similarly to those of NV centers. Therefore, we base our AC sensitivity calculations on a model developed for NV centers, evaluating both types of spin defects for dynamical decoupling AC measurement protocols (e.g., CPMG, XY8).  The AC magnetic field sensitivity for small signals in a regime dominated by shot-noise and/or spin projection noise is given by \cite{barry_sensitivity_2020}: 
\begin{multline}\label{eta}
\eta\approx\frac{\pi}{2}\frac{\hbar}{g_e \mu_B}\frac{1}{\sqrt{N \tau_{\text{full}}}} \frac{1}{e^{-\tau_{\text{full}}/(k^s T_2)^p}} \\ \times
\sqrt{1+\frac{1}{C^2 n_{\text{avg}}}}\sqrt{\frac{t_I+\tau_{\text{full}}+t_R}{\tau_{\text{full}}}}.
\end{multline}

In the above equation, $g_e$ is the g-factor of the relevant defect electronic spin ($V_B^-$or NV); $\mu_B$ is the Bohr magneton; $N$ is the number of defects in the sensing ensemble; $\tau_{\text{full}}$ is the full interrogation time; $T_2$ is the Hahn spin-echo coherence time of the spin defect; $k$ is the number of pulses in the optimized dynamical decoupling measurement protocol; $s$ is the power law scaling of the coherence time with pulse number $k$, which depends on the dominant decoherence source; $p$ is a stretched exponential parameter that characterizes the dominant source of defect spin dephasing \cite{bauch_ultralong_2018}; $C$ is the defect spin-state photoluminescence (PL) measurement contrast; $n_{\text{avg}}$ is the average number of photons collected per readout per defect; and $t_I$ and $t_R$ are the defect spin-state initialization and readout times, respectively. Note that $\eta$ $(\tau_{\text{full}} = 1\,\text{s})$ represents the smallest signal field that can be detected with SNR = 1 in 1 second of interrogation time. As several parameters in Eq.\ \ref{eta} depend on the frequency being sensed, we compare the performance of each system as a function of signal frequency. Defect NMR signals are typically on the order of MHz. 

Using Equation \ref{eta} as our framework, we make a series of modifications and additions to account for dynamics in an optimized experiment. First, we examine the dependence between variables, e.g., how an optimal $t_R$ varies with both $\tau_{\text{full}}$ and $t_I$. We model the PL measurement contrast, \textit{C}, by taking its average value over the course of the readout duration, with a decay profile defined by the base Rabi contrast and an exponential decay rate $1/t_I$. We optimize $t_R$ for every signal frequency by minimizing $\eta$ after accounting for the contrast decay profile. The best value is described by a transcendental equation accounting for the dependence of contrast, number of photons collected, and overhead time on $t_R$.  We implement a direct search optimization algorithm that solves for $t_R$ at every signal frequency (see Supplementary Note 2 and Supplementary Fig. 4 for more information).

 Next, we calculate \textit{k} using the formula for $k_{\text{opt}}$ defined in \cite{barry_sensitivity_2020}:
\begin{equation}
k_{\text{opt}} = \left[ \frac{1}{2p(1-s)} \left( \frac{2 T_2}{T_B} \right) ^p \right] ^{\frac{1}{p(1-s)}}
\end{equation}
where $T_B$ is the period of the AC signal field to be measured.  Since $T_B$ is inversely proportional to the signal frequency, $\omega$, and $k_{\text{opt}}$ has no range restrictions, $k_{\text{opt}}$ for extreme frequencies can result in numbers of $\pi$ pulses that are unrealistic to implement experimentally.  We thus restrict $k_{\text {opt}}$ to correspond to an integer number of (commonly used) XY8 measurement sequences, with a maximum value $(T_{2, \text{extended}}/T_{2, \text{echo}})^{1/s}$ such that $k^sT_2$ does not exceed the $T_2$ saturation value. 

In addition to considering the effect of dynamic parameters, we also account for the impact of finite pulse duration and the effectiveness of microwave (MW) driving for each defect system. Since the axes of quantization for NV centers and $V_B^-$ differ in orientation relative to the host material surface, we expect the Rabi frequency to be about $\sqrt{3}$ times faster for $V_B^-$ compared to NV centers given identical, experimentally typical values for the MW power and standoff distance between the antenna and defects. To model the impact of pulse duration, we subtract the time spent applying MW pulses to the defect spins from the interrogation duration $\tau_{\text{full}}$ after optimizing $k$ and other parameters.  The result of these simultaneous optimizations is a projection for AC sensitivity vs. frequency that indicates what can be expected from an optimized experiment. 

To compare the sensitivity of the two systems ($V_B^-$ and NV) for NMR measurements of samples on the material surface (hBN and diamond), we use parameters from three classes of state-of-the-art NV diamonds and two sets of $V_B^-$ parameters derived from our literature review. The three NV systems represent different use cases: a single NV in a nanopillar, a shallow NV ensemble, and a high-density NV ensemble uniformly distributed over a depth of $\sim10\,\mu$m  within the diamond (denoted as ``single NV'', ``shallow NV'' and ``bulk NV'', respectively). To assess the AC sensitivity of $V_B^-$ we use both (i) the parameters reported and derived from \cite{gao_high-contrast_2021} (``$V_B^-$ Gao'') and (ii) the best reported value of each parameter for $V_B^-$ across multiple publications (``$V_B^-$ Aggregated''). The latter set of parameters provides an estimate of what may be achievable in a single optimized experiment. In particular, \cite{gao_high-contrast_2021} does not implement isotopic enrichment of the hBN host or a plasmonic device such as a nanotrench antenna, which have been shown to increase $V_B^-$ spin coherence time and PL brightness, respectively \cite{gong_isotope_2024,cai_spin_2023}. All values used in our model are listed in Table \ref{TABLE}. See Supplemental Information for data used to inform our model calculations and discussion of derived values.

\begin{table*}
\begin{ruledtabular}
\begin{tabular}{|c|c|c|l|c|c|} \hline 
Parameter & \specialcell{"$V^{-}_B$ Gao" \\ Ref \cite{gao_high-contrast_2021} }& \specialcell{"$V^{-}_B$ Aggregated" \\ (Best Measured)}  &  Single NV \cite{qzabre_2025} &Shallow NV & Bulk NV \\
\hline
\hline
\specialcell{Echo $T_2$ \,($\mu$s)}& 1.1& 2\cite{gottscholl_room_2021} &  4&1.62& 10.7\\ \hline 
 Max $T_2$, dynamic decoupling ($\mu$m) & 4.4 \cite{ramsay_room_2022}& 4.4 \cite{ramsay_room_2022}& 50& 45.6&77\\
\hline
Depth (nm) & 2.5&  2.5&  10&10& 10-10,000\\ \hline 
 AC measurement PL Contrast& 4.25\%& 18\% \cite{lee_intrinsic_2025} & 27\%& 9\%&9\%\\
\hline
 Detected PL counts per defect (Hz)& 87.5$^\dagger$& 6000$^\dagger$\cite{cai_spin_2023}& 1,000,000& 50,000*&50,000*\\
\hline
Defect Density (ppm) & 192$^\dagger$& 236 \cite{gong_coherent_2023}&  NA&0.6& 2.7\\
\hline
Initialization Duration $t_i$ (ns)& 100 &  100&  2000 \cite{turner_quantum_nodate}&2000 \cite{turner_quantum_nodate}& 2000 \cite{turner_quantum_nodate}\\
\hline
Electron  g-factor& 2.001 \cite{gracheva_symmetry_2023}&  2.001 \cite{gracheva_symmetry_2023}&  2.003 \cite{barry_sensitivity_2020}&2.003 \cite{barry_sensitivity_2020}& 2.003 \cite{barry_sensitivity_2020}\\
\hline
Exponential scaling factor, $s$& 0.52 \cite{rizzato_extending_2023}&  0.52 \cite{rizzato_extending_2023}&  0.5*&0.58& 0.44\\ \hline
 Stretched exponential parameter, $p$& 1& 1& 2& 1&1\\ \hline
\end{tabular}
\end{ruledtabular}

\caption{\label{TABLE}Parameters used for calculations presented in this work. All unmarked parameters (no asterisk or dagger) represent experimentally measured values. Asterisks indicate estimated NV parameters, and daggers indicate estimated $V_B^-$ parameters (see Supplementary Note 1 and Supplementary Figs. 1-3 for supporting data for the NV systems and information about how marked $V_B^-$ parameters are determined). We consider a set of $V_B^-$ parameters from a single publication (reference \cite{gao_high-contrast_2021}) and optimistic set of parameters aggregated from multiple sources to account for uncertainty about $V_B^-$ properties and how optimization techniques can be combined. For NVs, we consider a single NV in a nanopillar, a shallow NV ensemble, and a 10$\,\mu$m thick, high density NV ensemble. }
\end{table*}

\begin{figure*}
    \centering
    \includegraphics[width=1\linewidth]{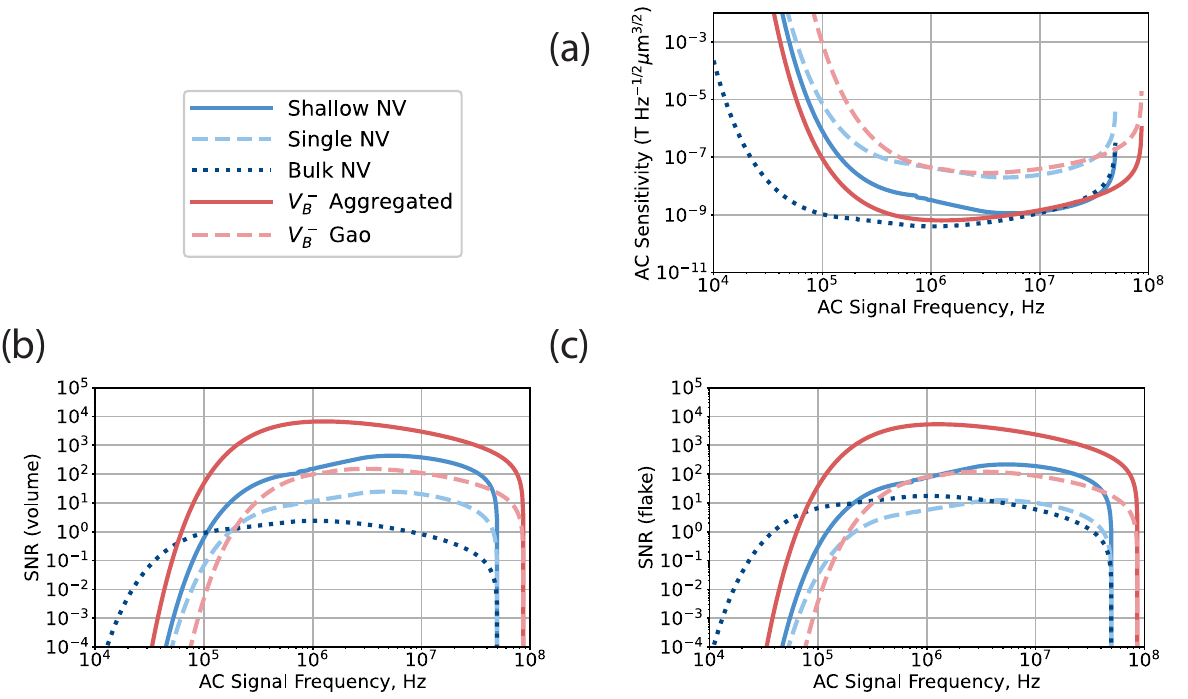}
    \caption{ \label{ac sens figure}(a) Volume-normalized AC sensitivity vs. AC signal frequency for three NV and two $V_B^-$ systems using dynamical decoupling detection; see Table \ref{TABLE}. Volume normalization for the ``Single NV'' system assumes 1 NV center per $\mu$m$^3$. AC sensitivity does not depend on standoff distance. Bulk NV systems excel in the low frequency regime due to their impressive coherence time $T_2$. At high frequencies, other measurement modalities perform well due to better PL brightness and contrast. Shallow NV systems have better spin coherence time and brightness per defect relative to $V_B^-$ systems; whereas $V_B^-$ exhibit higher defect density and PL contrast.  Sensitivity degrades for all systems at high frequency due to finite initialization and readout durations due to restrictions on the total number of pulses. (b) Calculated AC sensing SNR vs.\ signal frequency for a statistically-polarized, $\rho = 64\,\text{spins}/\text{nm}^3$, proton NMR signal occupying a semi-infinite cylinder above the material surface with 1 second of averaging time for different defect spin systems, assuming the standoff distance is equal to the defect depth. $V_B^-$ systems have a significant advantage due to their extremely shallow depths, requiring less sensitivity due to the stronger NMR signal, with $V_B^-$ Aggregated providing the best performance of any system over a wide frequency range. Bulk NV systems perform significantly worse than other modalities due to much larger mean standoff distance to the sample. (c) AC sensing SNR adjusted to accommodate a 1\,nm thin-layer sample on the material surface. Model predictions show that 2D samples are more sensitive to standoff distance than 3D samples, leading to increased performance of $V_B^-$ NMR systems.}
\end{figure*}

Using the approach described above, we estimate the volume-normalized sensitivity $\eta$ ($\tau_{\text{full}}$=1 s) for AC signals located at the material surface measured with dynamical decoupling measurement protocols as a function of detected signal frequency, with results shown in Fig. \ref{ac sens figure}a. As expected, the high-density, bulk NV ensemble exhibits the best sensitivity at lower signal frequencies. However, predicted SNR is reduced by contributions from NVs farther from the diamond surface and therefore the NMR signal source (Fig. \ref{ac sens figure}b). The $V_B^-$ parameters reported in \cite{gao_high-contrast_2021} (``$V_B^-$ Gao") result in worse sensitivity than optimized single NVs or shallow NV ensembles, but can still detect NMR signals with similar SNR to NV centers due to the smaller sensor-to-sample standoff distance. The best reported values for $V_B^-$ parameters (``$V_B^-$ Aggregated") yield sensitivity comparable to or better than single NV pillars and shallow NV ensembles; and the highest nanoscale NMR SNR of any system considered due to the combination of good sensitivity and shallow defect depth (i.e., small sensor/sample standoff).  Note that for all systems, signal averaging is incorporated into the model to resolve realistic nanoscale NMR signals with good SNR; hence the sensitivity values presented in Fig. \ref{ac sens figure}a (which represent just 1 second of averaging) do not directly correspond to the expected AC magnetic field amplitudes produced by nanoscale NMR signals. 

Notable behaviors arise from the analysis presented in Fig. \ref{ac sens figure}a.  First, sensor coherence time ($T_2$) has more relevance for low-frequency measurements than high-frequency measurements.  Specifically, long $T_2$ has almost no impact, in our model, on AC sensitivity for signal frequencies greater than 1\,MHz, since the exponential contribution $\text{exp}(\tau_{\text{full}}/(k^s T_2)^p)$ converges to 1 when $\omega$ is large (Supplementary Fig. 5).  This result helps explain why $V_B^-$ centers are able to maintain good AC sensing performance relative to bulk NVs at high frequencies: defect density, PL brightness and contrast, and achievable defect-spin Rabi frequencies play a dominant role in AC sensing performance in the high frequency regime. Although individual $V^{-}_B$  defects have lower PL brightness than NV centers, this difference is compensated by greater $V_B^-$ defect density and greater PL contrast for higher AC sensing frequencies.  

We also highlight the beneficial impact of shallow defect depth on sensitivity for nanoscale samples, which has inverse cubic scaling due to dipolar coupling between the defect sensor and the NMR signal source.  We evaluate this contribution in Fig.\ref{ac sens figure}b,  via the AC sensing SNR with 1 second averaging time for an NMR signal coming from a statistically polarized sample with proton density $\rho = 64$ nm$^{-3}$ located at the material surface, i.e., a few-nanometer-scale sample. We evaluate  $B_{rms}$, the root-mean-square AC magnetic field amplitude produced by the sample spins at the sensing plane for a given defect depth, as defined in \cite{pham_nmr_2016} and \cite{aslam_nanoscale_2017}.  Since $B_{rms}^2$ directly contributes to the AC measurement PL contrast for a statistically polarized sample, i.e., $C \sim$ exp$(-\gamma_e B_{rms}^2)$, changes in $B_{rms}^2$ dramatically impact the signal magnitude at the mean sensor location.  Large changes in  $B_{rms}^2$ and therefore $C$ can be accomplished by adjusting defect depth by only a few nanometers.  To quantify this effect, we calculate the measurement SNR from the ratio of $B_{rms}$ of a particular sample to the corresponding sensor sensitivity $\eta$ (see Supplementary Note 3 for further discussion): 
\begin{equation}
\label{SNR equation}
\text
{SNR} = \frac{(\mu_0 \hbar \gamma_n)}{32 \eta}\sqrt{\frac{G(\alpha)\rho}{2\pi d^3_r}}
\end{equation}
where $G(\alpha) = 8 - 3 \sin^4\alpha$ is a geometry scaling factor for $V_B^-$ and NV orientation defect angles, $\alpha$, relative to the material surface and NMR sample location (Fig. \ref{geometryfactor}); $\rho$ is the density sample spins in the sensing volume, $d_r$ represents the mean defect depth, and $\eta$ represents the $V_B^-$ and NV sensitivities to the desired AC signal with 1 second of averaging time. As many samples of interest consist of surfaces or two-dimensional materials, we account for samples of this form by adjusting Equation \ref{SNR equation} (see Supplementary Note 4). For this work we focus on NV centers in commonly available [110] cut diamonds. NV diamonds grown in the [111] direction have been described recently \cite{hughesStronglyInteractingTwodimensional2024a,wuSpinSqueezingEnsemble2025}, however shallow NV centers in such diamonds have not been characterized.

\section{Micron Scale Defect NMR}

\begin{figure*}
    \includegraphics[width=1\linewidth]{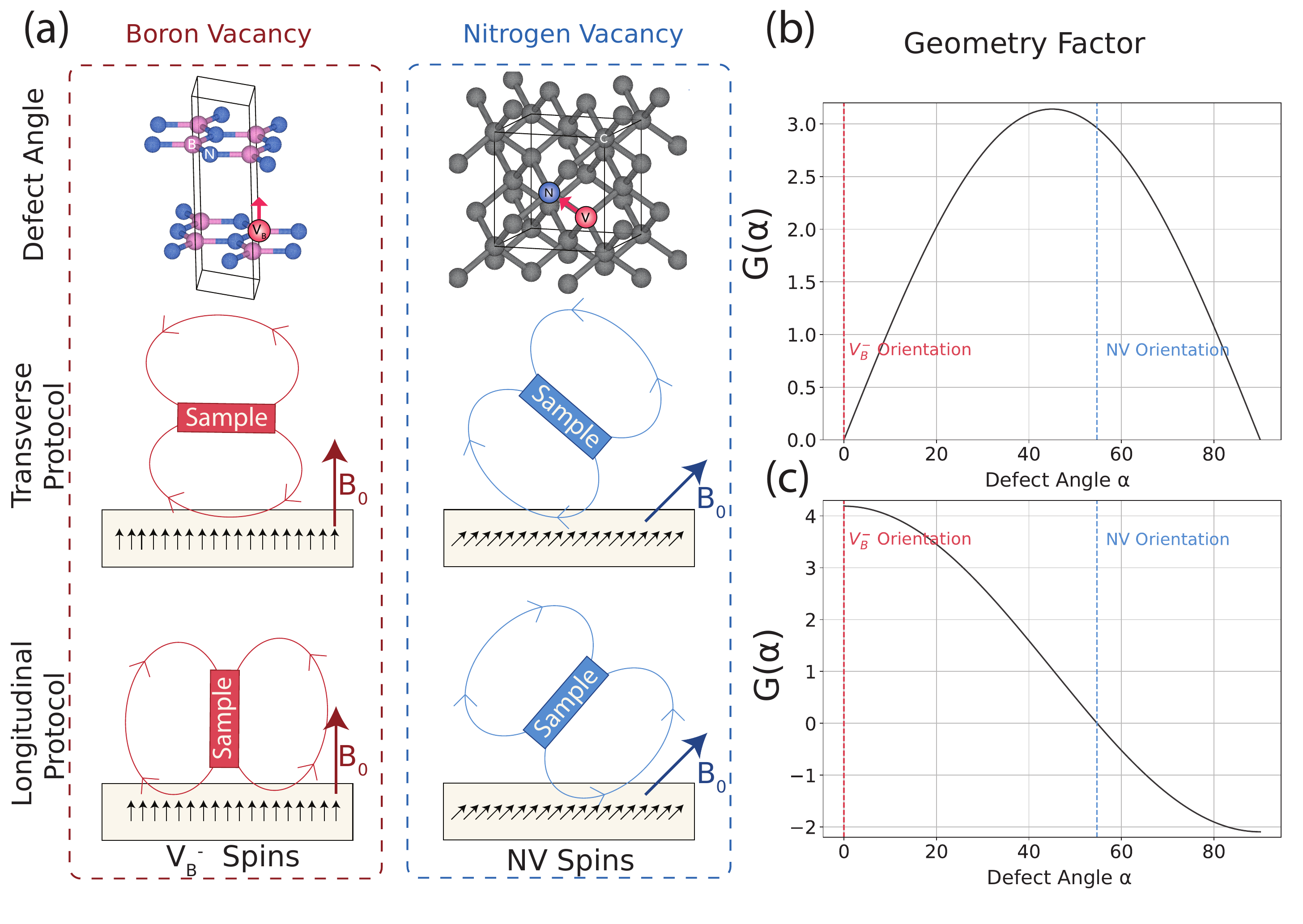}
    \caption{\label{geometryfactor}Geometry scaling factor calculations for transverse and longitudinal AC signals demonstrate the effect of defect angle $\alpha$ on signal phase accumulation efficiency.  (a) Schematics of ensemble $V_B^-$ and NV defects used for micron-scale sensing of transverse and longitudinal AC signals from samples near the material surface.  (b) Transverse AC signal indicative of an XY8-k or CASR measurement in micron-scale samples.  $V^{-}_B$ centers are completely insensitive with $G=0$; and NV centers, grown in the most common [110] orientation indicated by the vertical green line, are near-maximally sensitive.  (c) Longitudinal AC signal indicative of an AERIS or DRACAERIS measurement.  $V^{-}_B$ centers are now maximally sensitive, whereas NV centers have very low sensitivity.}
\end{figure*}

In defect NMR for sample volumes $>$1 $\mu$m$^3$, sample nuclear spins are typically thermally polarized \cite{glenn_high-resolution_2018} or hyperpolarized \cite{bucher_hyperpolarization-enhanced_2020, arunkumar_micron-scale_2021} at a bias magnetic field $\lesssim$ 0.1$\,$T, and both the sample nuclear spins and sensor electronic spins are coherently controlled with radiofrequency (RF) and MW pulse sequences, respectively \cite{aslam_quantum_2023}. As the bias field is typically aligned with the defect quantization axis, the sample longitudinal polarization direction is determined by the defect geometry (see Fig. \ref{geometryfactor}a).  In this micron-scale regime of uniform sample polarization (thermal or hyperpolarized), the orientation of the defect axis of quantization relative to the sample position and resulting NMR signal field direction significantly impacts the efficiency of dynamical phase accumulation by the defect sensor spins, which in turn affects the AC sensitivity.

Thermally polarized samples benefit from strong bias magnetic fields; however, the high-spectral resolution of Coherently Averaged Synchronized Readout (CASR) protocols using NV ensembles is only practical at sub-Tesla bias fields with sample NMR frequencies $\sim$ few MHz \cite{glenn_high-resolution_2018}.  CASR circumvents limits to spectral resolution set by the finite NV $T_2$ by performing sequential $T_2$-limited NV-NMR measurements synchronized to a clock and reconstructing the $T_2$-unlimited NMR signal from the aliased form of the sequential measurements. At tesla-scale bias fields with NMR signal frequencies $\sim$ 100 MHz, it becomes impractical to apply sufficiently strong and fast MW pulses to the NV electronic spins for resolvable NMR signal detection. Instead, longitudinal NMR detection sequences such as AERIS \cite{munuera-javaloy_high-resolution_2023,whaitesHighFieldNMREnhanced2025} or DRACAERIS \cite{daly_nutation-based_2024} have been proposed to downconvert the driven NMR signal, enabling practical NV measurement at frequencies $\sim$ few MHz. Also, quantum frequency mixing (QFM) may allow application of CASR to NV-NMR samples at mutli-tesla bias fields, albeit with reduced sensitivity \cite{yin_high-resolution_2024}.

$V^{-}_B$ defects are not well suited to perform CASR in the micron-scale regime due to the geometric orientation of the defect axis of quantization, as shown in Fig. \ref{geometryfactor}b. (Note that the nanoscale regime with dominant statistical polarization has different geometry considerations -- see Supplementary Fig. 6). A recent $V_B^-$ CASR demonstration \cite{rizzato_extending_2023} used test signals oriented orthogonal to those created by a micron-scale $V_B^-$ NMR experiment.

In contrast, $V^{-}_B$ ensembles have a large geometric advantage relative to NV ensembles for longitudinal NMR detection, as the $V_B^-$ axis of quantization is aligned with the maximally sensitive orientation for longitudinal protocols (see Fig. \ref{geometryfactor}c).  Following the model used in \cite{bruckmaier_geometry_2021}, we calculate the geometry factor for ensemble $V_B^-$ detection of longitudinal NMR signals to be $G = \frac{4}{3}\pi$, whereas for NVs $G$ is nearly zero (see Supplementary Note 3).  

\section{Viability of Statistical Polarization $V_B^-$ NMR}

For nanoscale NMR samples, statistical polarization is typically much larger than either thermal polarization or hyperpolarization \cite{herzog_boundary_2014}. To assess the viability of nanoscale $V_B^-$ NMR on statistically-polarized samples, we assume an XYn-k experimental protocol resembling the technique used in \cite{pham_nmr_2016} as a framework for our simulations.  XYn-k protocols operate by performing \textit{k} repeated XYn subsequences on defect spins followed by optical readout. The frequency sensitivity is determined by the free precession interval, $\tau$; hence by performing measurements for a series of values for $\tau$, a spectrum of the sample NMR signal can be reconstructed \cite{pham_nmr_2016}.  Without a signal present, the XYn-k measurement contrast exponentially decays with the defect spin coherence time $T_2$ \cite{rizzato_extending_2023}; introduction of an AC signal produces a dip in the decaying XYn-k PL measurement contrast. Well-established data analysis techniques \cite{mamin_nanoscale_2013,staudacher_nuclear_2013,lovchinsky_nuclear_2016,glenn_high-resolution_2018,pham_nmr_2016,aslam_nanoscale_2017,bruckmaier_imaging_2023,bucher_hyperpolarization-enhanced_2020,schwartz_blueprint_2019} then identify the AC Fourier components of the signals of interest, e.g., the NMR spectrum from a nanoscale sample near the $V_B^-$ sensor.  XYn-k protocols also form the basis for defect NMR measurement schemes with high spectral resolution, such as Correlation Spectroscopy \cite{liu_surface_2022} and CASR \cite{glenn_high-resolution_2018}.

Using this framework, we evaluate the performance of $V^{-}_B$ NMR in the statistical polarization regime compared to established NV-NMR techniques. Note that boron and nitrogen nuclear spins in the hBN lattice are not considered here as a source of background $V_B^-$ NMR signal, as their gyromagnetic ratios are very different from that of typical sample target nuclear spins, such as $^1H$, $^{13}C$, $^{2}H$,$^{19}F$, and $^{31}P$. 
 
\subsection{Sensor-Sample Back-Action}

\begin{figure*}
 \includegraphics[width=1\linewidth]{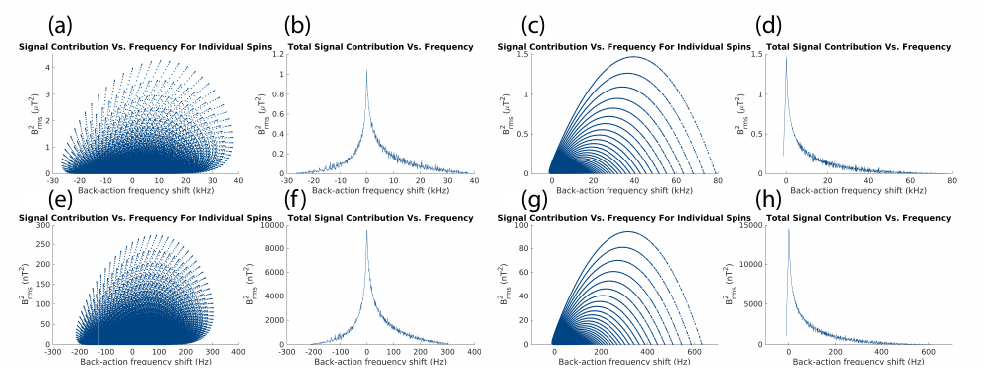}
    \caption{\label{BackActionBroadening}Model calculations of differences between NV and $V_B^-$ NMR properties due to sensor-sample back-action. In the model calculations, protons of polarized spin density $\rho = 64 \text{ spins/nm}^3$ are confined to a $(4d)^3$ hemispherical volume above the surface, where $d$ is the sensor depth. (a) Back-action shift of each sample spin caused by a $d=1$\,nm deep NV electronic spin.  (b) Total NMR signal measured by the NV sensor, which exhibits homogeneous NMR lineshape broadening of $\sim$2 kHz. (c) Back-action shift of each sample spin caused by a $d=1$\,nm deep $V_B^-$ electronic spin. (d) Due to the out of plane $V_B^-$ angle, the NMR signal in the presence of a single $V_B^-$ defect exhibits inhomogeneous NMR lineshape broadening of $\sim$2 kHz. (e) Back-action shift of each sample spin caused by a $d=5$\,nm deep NV electronic spin.  (f) Total NMR signal measured by the NV sensor, which exhibits homogeneous NMR lineshape broadening of $\sim$25 Hz. (g) Back-action shift of each sample spin caused by a $d=5$\,nm deep $V_B^-$ electronic spin.  (h) Total NMR signal measured by the $V_B^-$ sensor, which exhibits inhomogeneous lineshape broadening of $\sim$20 Hz.}
\end{figure*}

\begin{figure}
 \includegraphics[width=1\linewidth]{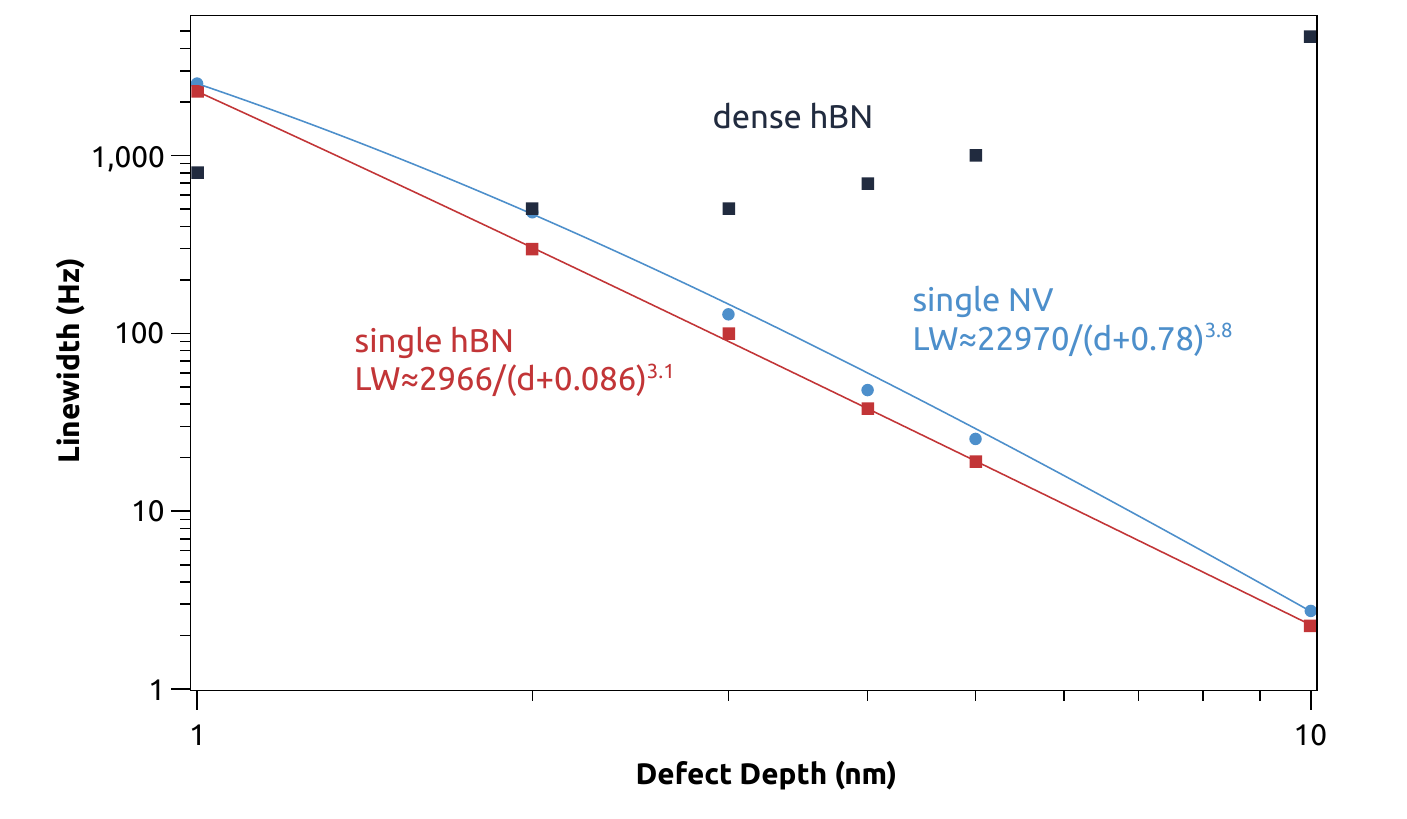}
    \caption{\label{BackActionBroadening2}Simulated NMR linewidth scales as a power law with defect depth $d$ for both single NV (blue) and single $V_B^-$ (red) systems. For dense $V_B^-$ layers, the linewidth remains approximately constant (black). See Supplementary Fig.\ 7 for additional data on dense hBN ensembles.} 
\end{figure}

At small distances, sensor-sample back-action can perturb the resonance frequency of both the sample and defect spins. The effect arises from the magnetic dipolar interaction, which for a single defect and a single sample spin is given by
\begin{equation}
\label{smallensemblebackaction}
\hat{H}_D = \frac{\mu_0}{4 \pi} \gamma_I \gamma_{S} \hbar \left( \frac{1}{r^3} \left[ 3 (\hat{r} \cdot \hat{I})(\hat{r} \cdot \hat{S}) - \hat{I} \cdot \hat{S} \right] \right),
\end{equation}
where $\hat{I}$ and $\hat{S}$ are the spin angular momentum operators for the sample and defect spins, respectively, $\hat{r}$ is the unit vector defining the direction between the two spins, and $r$ is the distance between them. For measurements using single defects or a low-density of defects, only the back-action from the sensing defect will have an effect, since the $1/r^3$ dependence means defects more than $\sim$10 nm away have little influence. However, when defect density is dense, each sample spin can couple to multiple defects, leading to additional back-action effects.

We simulate the effect of quantum mechanical back-action modifying the sensor and sample spin resonance properties. For both NV and $V_B^-$, we model the NMR signal contributions from a polarized spin density $\rho = 64 \text{ spins/nm}^3$ above the sensor surface, sampling up to 28 million sample spin locations. The calculated NMR signal broadening differs between NV and $V_B^-$ systems as a result of their different orientations (Fig. \ref{BackActionBroadening}). Since the $V_B^-$ orientation is perpendicular to the surface, the closest proton spins all exhibit coupling to the $V_B^-$ defect of the same sign, and thus to an inhomogeneous NMR lineshape. For the shallow NV sample, the lateral NV-NV distance is $\sim 17$ nm, and back-action on a given proton spin from neighboring NVs is insignificant. For ten layers of hBN, the lateral inter-defect distance is $\sim 1.4$ nm, and there is large back-action on a given proton spin from neighboring $V_B^-$ defects, leading to NMR lineshape broadening that is relatively independent of defect depth (see Supplemental Information). For individual defects, NMR lineshape broadening effects is slightly less severe for $V_B^-$ NMR than NV-NMR (Fig. \ref{BackActionBroadening2}), but dense $V_B^-$ is only better at the smallest depths. Note that our simulation assumes sensor spin states are in the superposition used during sensing (i.e. $\psi = (\ket{+1}+\ket{0})/2$) and is half the maximum possible value (when $\psi = \ket{+1}$). Broadening would not be present when the sensor is in the $\psi = \ket{0}$ state. It should also be noted that the dipolar interaction affects the sensor electronic spin resonances, which will be shifted by the mean of the back-action frequency shift. For NVs, the mean is near zero, but for dense $V_B^-$ the shift can be 10 kHz or more. See Supplementary Note 5 for additional back-action simulations. 

Previous work with NVs \cite{childress_coherent_2006,taminiau_detection_2012} provides a possible approach to exploit back-action and allow $V_B^-$ sensors to selectively measure the NMR spectra from individual nearby ($\sim$1 nm) sample nuclear spins.  In these NV experiments, back-action from a single NV electronic spin is used to manipulate nearby (within a few nm) $^{13}$C nuclear spins in the diamond host material, allowing individual $^{13}$C spins to be selectively detected \cite{childress_coherent_2006,taminiau_detection_2012}. Dynamical decoupling techniques are used to distinguish proximal nuclear spins from the spin bath, allowing for the measurement of both the three-dimensional structure of a nuclear spin cluster and the position of the NV sensor relative to the cluster. However, application of this approach to NMR measurements of individual nuclear spins outside of the diamond is limited by the depth of the NV defect \cite{abobeih_atomic-scale_2019}. $V_B^-$ NMR may be able to build on the advances from NV-NMR in nearby nuclear spin control while taking advantage of reduced standoff distances to nuclear spins on the hBN surface \cite{childress_coherent_2006, taminiau_detection_2012, abobeih_atomic-scale_2019, bradley_ten-qubit_2019, sushkov_magnetic_2014, lovchinsky_nuclear_2016, abobeih_one-second_2018, dutt_quantum_2007}.

\subsection{Effects of Sample Diffusion}

\begin{figure}
    \includegraphics[width=1\linewidth]{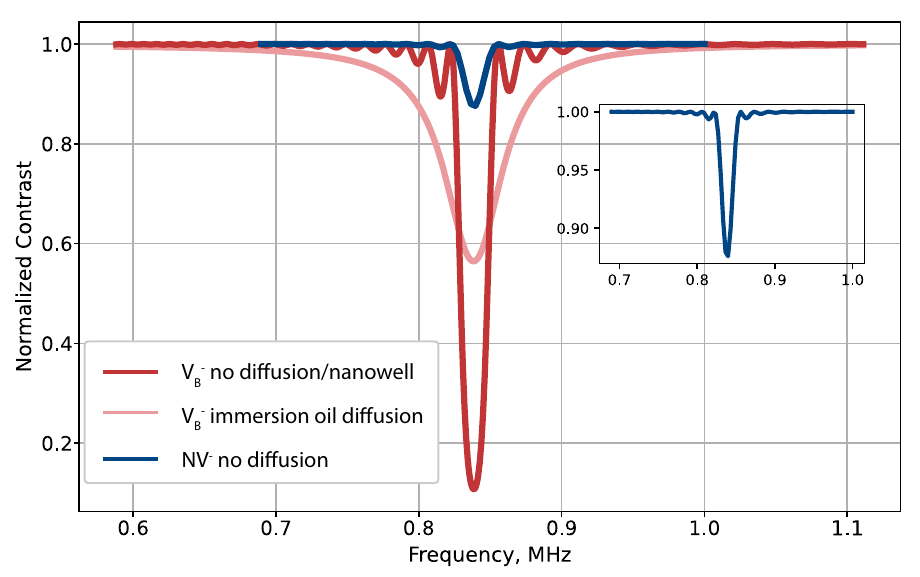}
    \caption{\label{ContrastComparison}Model calculation of nanoscale NMR lineshapes for a XY8-100 measurement pulse sequence and single $V^{-}_B$ and NV sensors.  When measuring low-concentration samples, the shallower $V_B^-$ defect gives larger NMR signal contrast than the deeper NV center. The modeled NMR signal is from a sample containing $^1$H (proton) spins at $B_z$ = 0.0197\,T, with density $\rho = 1$ spins/nm$^3$, $d_{V^{-}_B} = 2.5$\,nm, $d_{NV} = 6$\,nm, and defect angle $\alpha$=0 for the $V_B^-$ sensor to indicate out-of-plane orientation.  Sample spin dephasing time $T_2^*$ is taken to be infinite for the no diffusion case. For the hBN nanowell, D is experimentally measured to be $\approx 118\,$s\cite{clark_tracking_2022}, which can be neglected. For finite sample diffusion, $T_{2,N}^{*}$ is limited by the correlation time function $T_D(D),$ where $D$ is the sample diffusion coefficient, with $T_{2,N}^{*} \sim 9\times10^{-6}$\,s for immersion oil. \color{black} The inset shows a zoomed-in version of the NV-NMR spectrum in the absence of diffusion.}
\end{figure}

To study the impact of sample diffusion on nanoscale NMR sensitivity and lineshape, including differences between $V^{-}_B$ and NV sensors, we model the contrast function $C(\tau)$ defined by \cite{pham_nmr_2016} for an XY8-64 experiment measuring the statistically-polarized NMR signal from a sample with a freely-diffusing nuclear spin density $\rho = 2.95\,\text{nm}^{-3}$; and compare the NMR signal measured by a single 6\,nm deep NV and 2.5\,nm deep $V^{-}_B$ sensor (Fig. \ref{ContrastComparison}).  See the Supplemental Information for further details. Notably, we find that the $V^{-}_B$ system exhibits $\sim$16x greater measurement contrast compared to the NV experiment, primarily due to the favorable spin depth and geometry of the $V^-_B$ defect, which boosts $B_{rms}$ by power-law scaling (see Fig. \ref{geometryfactor}c).  In these model calculations, both $V^{-}_B$ and NV measurements have detection-bandwidth-limited NMR linewidths of approximately 100\,kHz. The detection bandwidth can be narrowed by increasing the number of $\pi$ pulses, \textit{k}, limited by decoherence of the defect spin sensor; to $\sim$ 1 kHz using correlation spectroscopy (sensor spin $T_1$ limit); and to $\sim$ 1 Hz with CASR \cite{glenn_high-resolution_2018}. 

At the nanoscale, diffusion of liquid samples causes large spectral broadening in NMR measurements, which can greatly limit utility for chemical identification \cite{claridge_chapter_2009}. This phenomenon is a result of the limited interaction time between sample spins and defect sensor spins due to molecular diffusion. The diffusion-induced decay of the sample magnetization measurable by the sensor spin is characterized by the spin correlation time $T_D$ to lowest order, which can be estimated assuming a 3D random walk \cite{chandler_introduction_1987}:
\begin{equation}
T_D \approx d^2/6D
\end{equation}
where \textit{d} is the defect sensor distance from the sample and \textit{D} is the diffusion coefficient of the sample.  Nuclear spin dephasing $T^{*}_{2, N}$ is restricted by the correlation time $T_D$; and the measurement contrast $C(\tau)$ directly depends on \textit{k}, $\tau$, and $T^{*}_{2, N}$ in the low $T^{*}_{2, N}$ limit. Fig. \ref{ContrastComparison} also shows an example of the calculated effects of diffusion broadening on nanoscale $V_B^-$ NMR spectra using an XY8-100 protocol.

To limit diffusion broadening in nanoscale $V_B^-$ NMR experiments, a liquid sample can be confined in nanowells etched into hBN layers using electron beam lithography.  Encapsulating hBN layers containing $V_B^-$ defects could be placed above and below nanowells in place of the graphene layers used in \cite{clark_tracking_2022}.  PL from unwanted areas of a homogeneously irradiated hBN layer (i.e., not near a nanowell) could be obscured using nanolithography, with few nm precision \cite{sharma_evolution_2022}.  Alternatively, $V_B^-$ sensors could be preferentially created in an encapsulating hBN layer, near each nanowell, with 10 nm precision, using a focused ion beam (FIB)  \cite{gierak_focused_1997,kianinia_generation_2020}. Diffusion in nanowells is greatly reduced compared to typical values for macroscopic liquid samples such as immersion oil; see Fig. \ref{ContrastComparison}. For example, by tracking the motion of individual platinum ions dissolved in water and contained within nanowells of 30\,nm height and 100 nm radius, past work \cite{clark_tracking_2022} showed that the effective diffusion coefficient is reduced from the bulk value (D $\sim5\times10^5 \,\text{nm}^2s^{-1}$) to $0.038\,\text{nm}^2 \text{s}^{-1}$. See also the modeling of nanowell confinement for NV-NMR \cite{cohen_confined_2020}, which reaches a similar conclusion. 

\section{Discussion}

Boron vacancy ($V_B^-$) defects in hBN may provide new or enhanced capabilities in nanoscale and micron-scale (ultralow-mass) NMR. The model calculations outlined above indicate that  $V^{-}_B$ AC sensitivity can be comparable to that of shallow NVs at MHz signal frequencies. In the nanoscale regime of statistically polarized nuclear spin samples, the typically smaller standoff distance of $V_B^-$ defects relative to NVs provides a significant benefit to measurement contrast and SNR, which could surpass NV performance by over an order of magnitude. These results may be surprising given the relatively poor DC sensitivity and coherence times of $V_B^-$ defects, indicating that $V_B^-$ AC sensing applications are potentially underappreciated. In micron-scale samples, where uniform polarization (thermal or hyperpolarized) is typically dominant, $V^{-}_B$ is well-suited for high-field (tesla-scale) applications using longitudinal NMR detection protocols, due to geometric considerations.  At the near-micron-scale ($\sim$300$\,$nm), statistical polarization can be greater than thermal polarization, depending on details of the sample and experimental set-up \cite{herzog_boundary_2014, glenn_high-resolution_2018}, allowing defect NMR measurements with diffraction-limited optics and a relatively simple apparatus, e.g., using an ensemble of $V_B^-$ in a homogeneously irradiated hBN flake.

For low-concentration NMR samples, the high density of $V^{-}_B$ defects enhances AC sensing SNR compared to NVs. In combination with shallow depths, $V^{-}_B$ defects may be well-suited for studies of ultralow-mass and mass-limited samples, with potential applications, e.g., in pharmaceutical development where producing larger quantities can be resource intensive. The sensitivity of $V_B^-$ NMR may also facilitate the study of interface dynamics currently out of reach of conventional NMR techniques. 

Additionally, future work on the back-action between $V_B^-$ electronic and target nuclear spins on the hBN surface or in a nanowell may provide new methods for atomic-scale measurement of individual spins within the sample. With NVs, similar strong electronic-nuclear dipolar coupling has been used successfully for atomic-scale localization and measurement of nuclei within the diamond \cite{abobeih_atomic-scale_2019}, but to date has not been practical for samples outside the diamond host. 

$V^{-}_B$ NMR can also leverage other functional roles of hBN.  Natural targets for $V_B^-$ NMR include the study of liquid crystal alkanes on the hBN surface \cite{palinkas_composition_2022}, or of materials typically encapsulated by hBN such as 2D heterostructures or thin film perovskites \cite{novoselov2DMaterialsVan2016a, zhaoEnhancedPerformanceLongterm2025,seitz_long-term_2019}. hBN is well-suited for nanowell fabrication, which can effectively eliminate diffusion broadening of NMR spectra at the nanoscale \cite{clark_tracking_2022}. Such developments are likely critical for controlling diffusion broadening and enabling high spectral resolution NMR of nanoscale liquid samples using shallow defect sensors.

To develop useful $V^{-}_B$ NMR, future work will need to improve the standardization and repeatability of $V_B^-$ and hBN properties as well as optimize instrumental design and measurement protocols. For example, the measured $V_B^-$ spin coherence time ($T_2$) varies considerably with unknown cause in published measurements to date; understanding the mechanism of this variability and reliably producing hBN samples with longer coherence times will be of significant value to $V_B^-$ sensing applications.  Work is also needed to improve the consistency of fabricating hBN samples with specific thickness and $V_B^-$ defect depth. An optimized $V_B^-$ NMR experiment may require integrating irradiated hBN flakes into a nanowell structure and a gold nanotrench array (NTA) for enhanced MW driving of defect electronic spins, which will likely require careful parameter optimization, as NTA behavior is influenced by the thickness of the adjacent material \cite{cai_spin_2023}.  Additional benefits may come from optimizing $V_B^-$ sensor properties, e.g., by manipulating charge state, using ultra-pure hBN, and optimizing the optical excitation wavelength. 

Advances in data processing of defect measurements may also be required. Projected $V_B^-$ NMR spectra carry rich spatial information about sample spins, but in larger systems it may prove difficult to back out the desired information from a given spectrum, especially with ensemble-based measurements.  Developments in data analysis and algorithmic or machine learning-assisted interpretation of measurements may be used to sort and recover critical scientific information as $V^{-}_B$ NMR gains traction.

\section{Data Availability}

Experimental data to support the NV parameters used in this paper (Supplemental Figs. 2 and 3) is available at \url{https://doi.org/10.7910/DVN/ABGNZB} \cite{datacode_repository}. 

\section{Code Availability}

Code used to generate all simulations is available at \url{https://doi.org/10.7910/DVN/ABGNZB}  \cite{datacode_repository}.

\section{Author Contribution Statement}

*Denotes equal contribution between D.M.D. and N.R.R. 

J.W.B. conceived the project. D.M.D. and N.R.R. developed the idea, planned the analysis and directed the strategic vision. N.R.R. performed the literature analysis; developed the framework for analysis, and gathered parameters used in the simulations.  D.M.D. and S.J.D. performed all simulations and numerical calculations. Z.Y., J.C., and S.J.D. contributed to experimental analysis of NV samples and informed the parameter choices of NV samples. J.W.B., S.J.D., R.L.W. informed the strategic direction and supervised the project. D.M.D. and N.R.R. wrote the manuscript. All authors contributed to ideation and technical background, discussed the results and implications, and reviewed the manuscript. 

\section{Acknowledgments}

This material is based on work by N.R.R. supported by the National Science Foundation Graduate Research Fellowship Program under Grant No. DGE 2236417. This work is supported by, or in part by, the U.S. Army Research Laboratory under Contract No. W911NF2420143; the U.S. Army Research Office under Grant No. W911NF2120110; the U.S. Air Force Office of Scientific Research under Grant No. FA9550-22-1-0312; the Maryland Procurement Office under Award No. H9823019C0220; the Gordon \& Betty Moore Foundation under Grant No. 7797.01; and the University of Maryland Quantum Technology Center.

\section{Competing Interest Statement}

The authors declare no competing interests. 

\clearpage
\renewcommand{\thefigure}{S\arabic{figure}}
\setcounter{figure}{0}
\renewcommand{\thetable}{S\arabic{table}}
\setcounter{table}{0}
\renewcommand{\theequation}{S\arabic{equation}}
\setcounter{equation}{0}
\renewcommand{\thesection}{Supplementary Note \arabic{section}}
\setcounter{section}{0}

\section{Sensing Parameters}

\subsection{Estimating $V_B^-$ PL Brightness}

\begin{figure*}[h]
    
    \includegraphics[width=1\linewidth]{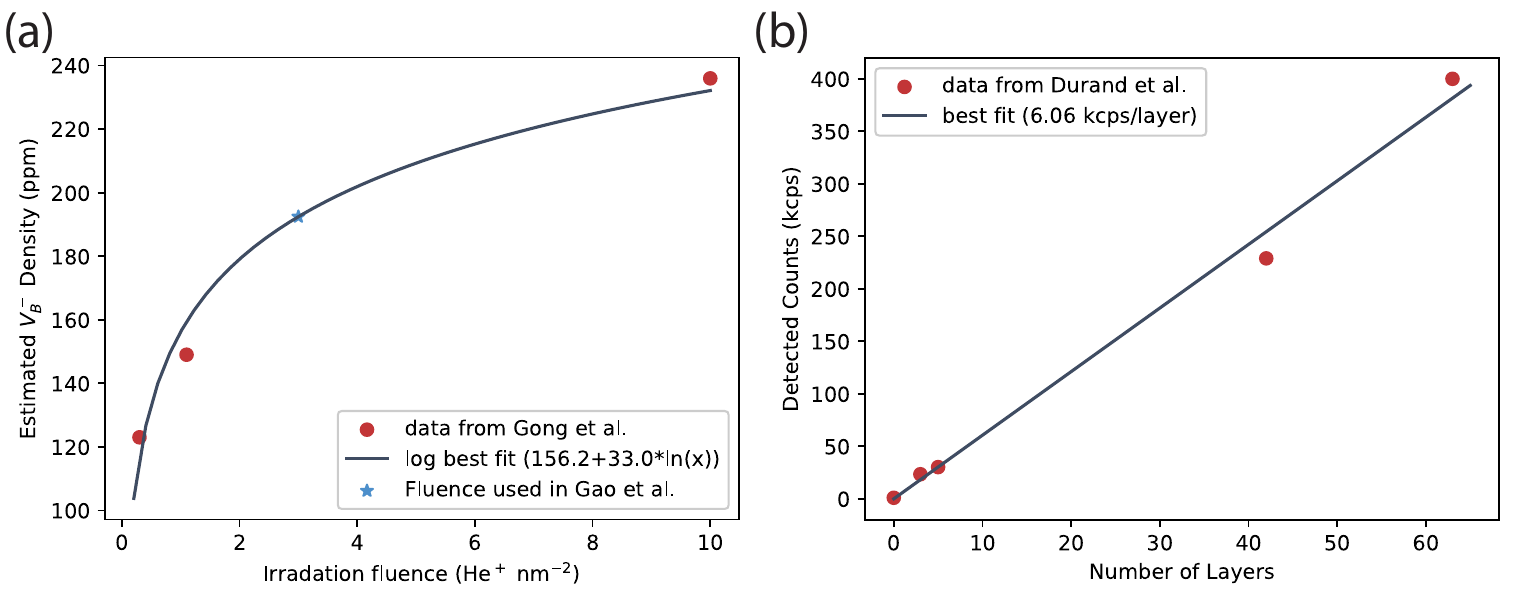}
    \caption{\label{brightness} Results from past work used in $V_B^-$ PL brightness/density estimations. (a) Estimated $V_B^-$ density vs He+ irradiation fluence from \cite{gong_coherent_2023}. Fit and interpolation provides estimate of $V_B^-$ density for samples described in \cite{gao_high-contrast_2021} (blue star). (b) Brightness (detected PL counts) vs estimated hBN layers from \cite{durand_optically_2023}, with a fit yielding an estimate of $V_B^-$ ensemble brightness per layer. }

\end{figure*}

As isolated $V_B^-$ defects have not been reported, we do not have direct measurements of single emitter photoluminescence (PL) brightness.  However,  the total number of PL photons detected across a $V_B^-$ ensemble should depend on the density of emitters, the interrogation volume, and the single emitter brightness. We therefore constrain $V_B^-$ density and brightness estimations to ensure they reproduce the observed ensemble PL brightness given the spot size and flake thickness of the proposed experiments.

We estimate the $V_B^-$ density of the sample described in \cite{gao_high-contrast_2021} by interpolating the reported $V_B^-$ density vs.\ fluence data from \cite{gong_coherent_2023}. Both works used He\textsuperscript{+} irradiation to create vacancies (and thus $V_B^-$ defects), with \cite{gao_high-contrast_2021} using  2.5 keV He\textsuperscript{+} ions for irradiation at a dose of 3 He\textsuperscript{+} nm$^{-2}$ and \cite{gong_coherent_2023} using 3 keV He\textsuperscript{+} ions, at 0.3, 1.1, and 10 He\textsuperscript{+} nm$^{-2}$. We use a log fit for the interpolation, and estimate that a fluence of 3 He\textsuperscript{+} nm$^{-2}$ would produce a $V_B^-$ density of 192 ppm.

To estimate the brightness of the emitters in \cite{gao_high-contrast_2021}, we calculate the diameter of the spot size based on the reported NA (0.9) and wavelength (532 nm) using the approximation $ d=\frac{\lambda}{2NA}$. This calculation results in an estimated spot size of 295.5$\,$nm, which corresponds to 485 $V_B^-$ per layer given a defect density of 192 ppm. Based on SRIM simulations showing defects concentrated in a 25 nm depth range (which would contain $\sim$ 75 hBN layers), we estimate each defect contributes about 88 cps to the final PL measurement. 

We also estimate the single defect brightness based on the results provided in \cite{durand_optically_2023}.  The authors report PL brightness for flakes believed to be 3 and 5 layers thick, as well as 15 and 22 nm flakes. As collection efficiency is approximately linear below 20 nm \cite{clua-provost_impact_2024}, we perform a linear fit enforcing the reported background of 1 kcps, resulting in an estimated per layer PL brightness of about 6.1 kcps.  It is more difficult to estimate the emitter density of these samples as they were produced through neutron irradiation instead of helium irradiation. However, if we assume a defect density of 192 ppm as before, and adjust for the 0.8 NA objective, we estimate each defect contributes 9.4 cps to the measured brightness.  This estimate is an order of magntiude lower than the value for \cite{gao_high-contrast_2021} discussed above, which is consistent with the plasmonic enhancement from the gold stripline used in \cite{gao_high-contrast_2021}. 

Note that the density of hBN is not isotropic, as the spacing between lattice sites within each layer is substantially smaller than the inter-layer spacing \cite{pease_crystal_1950}. For the purpose of the volume-normalized AC sensitivity estimates presented in Figure 3 of the main text, we assume an atomic density (and therefore $V_B^-$ density in ppm) based on the bulk properties of hBN in a 1$\,\mu$m cube.  However, it is worth noting that the actual number of $V_B^-$ defects in a few layer hBN sample with the same volume could be up to a factor of 2 larger. We use a model of the hBN lattice and the parameters reported in each paper to correct for this effect in our PL brightness estimations above. The increased number of $V_B^-$ defects in few-layer hBN may lead to a slight improvement ($\sim\sqrt{2)}$) of the AC sensitivity estimates shown in Figure 3.

For the ``$V_B^-$ Aggregated'' brightness (``Detected counts per defect'') listed in Table II of the main text, we begin with the unenhanced brightness of 9.4 cps/defect estimated from \cite{durand_optically_2023}, and assume a plasmonic enhancement of about 600$\times$. To date, local plasmonic-enhancement of $V_B^-$ spin defects up to 1600$\times$ has been reported \cite{xu_greatly_2023}, as well as enhancement of at least 400$\times$ across a micron-scale device \cite{cai_spin_2023}.

\subsection{Experimental Measurements of NV parameters}

\subsubsection{``Bulk NV" Parameters}

The bulk NV diamond sample used in this work is a high-purity ($>99.99\% $ $^{12}$C), CVD-grown diamond plate from Element Six.  The diamond measures ($2 \times 2 \times 0.5$)-mm$^3$, is side polished, and contains a 10$\,\mu$m thick nitrogen-15 rich layer grown along the [110] crystal orientation. Post-growth treatment through electron irradiation and annealing increases the concentration (density) of negatively charged NV centers to about 2.7$\,$ppm.

\begin{figure*}[h]
    \includegraphics[width=\linewidth]{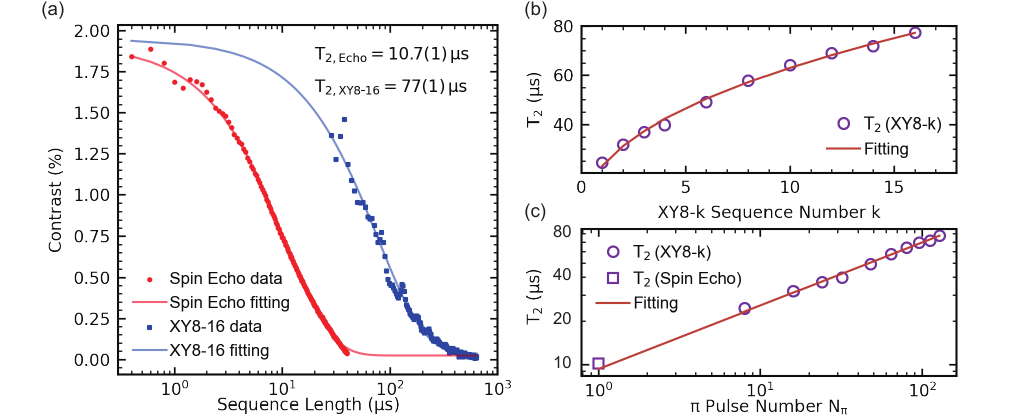}
    \caption{
    NV ensemble spin coherence time $T_2$ data and fitting results for ``Bulk NV" parameters listed in Table II of main text. (a) $T_2$ measured using Hahn Spin Echo (red) and XY8-16 (blue) sequences. Exponential fitting determines $T_{2,Echo}$ = 10.7(1)$\,\mu$s and $T_{2,XY8-16}$=77(1)$\,\mu$s. (b) $T_2$ measured for varying XY8-$k$ sequence number $k$, with associated fit.  $T_2$ is extended with increasing XY8 sequence numbers, reaching about  80$\,\mu$s for $k=16$. (c) Linear fit of $T_2$ with respect to $\pi$ pulse number $N_\pi$ in the dynamical decoupling sequence on a log-log scale. The fitted slope is $s$=0.44(1).
    }

    \label{fig:8}
\end{figure*}

To measure the NV spin coherence time $T_2$, we first perform a Hahn spin-echo experiment, yielding $T_{\mathrm{2,Echo}} = 10.7(1)\,\mathrm{\mu s}$, as shown in Figure \ref{fig:8}a. Subsequently, we measure $T_2$ using XY8-$k$ sequences with varying $k$. The coherence time increases with the number of sequences $k$, reaching  $T_{\mathrm{2,max}} \approx 80\,\mathrm{\mu s}$ for $k=16$, as illustrated in Figure \ref{fig:8}b.

As the number of $\pi$ pulses, $N_\pi$, in the dynamical decoupling sequence increases, the coherence time exhibits a power-law dependency, $T_2 \propto T_{2,Echo} N_\pi^s$ where $s$ is the exponential scaling factor in Equation 2 and Table II of the main text. To extract $s$, we fit the data linearly on a log scale, as shown in Figure \ref{fig:8}(c), yielding $s=0.44(1)$. This dataset is also available in the Appendix of \cite{yin2024}.

The difference between the AC measurement PL contrast demonstrated in Figure \ref{fig:8}(a) and the $9\%$ value reported in Table II of the main text arises due to different normalization schemes. For the spin-echo and XY8-16 measurements, we measure two bulk NV PL signals, $S_1$ and $S_2$, with a $180^\circ$ phase difference in the last $\pi/2$ pulse. The signal is normalized as $(S_1-S_2)/(S_1+S_2)$. The $9\%$ value in Table II comes from a generalized contrast calculation defined as signal/reference (no MW pulses).

\subsubsection{``Shallow NV'' Parameters}

\begin{figure*}[h]

    \includegraphics[width=\linewidth]{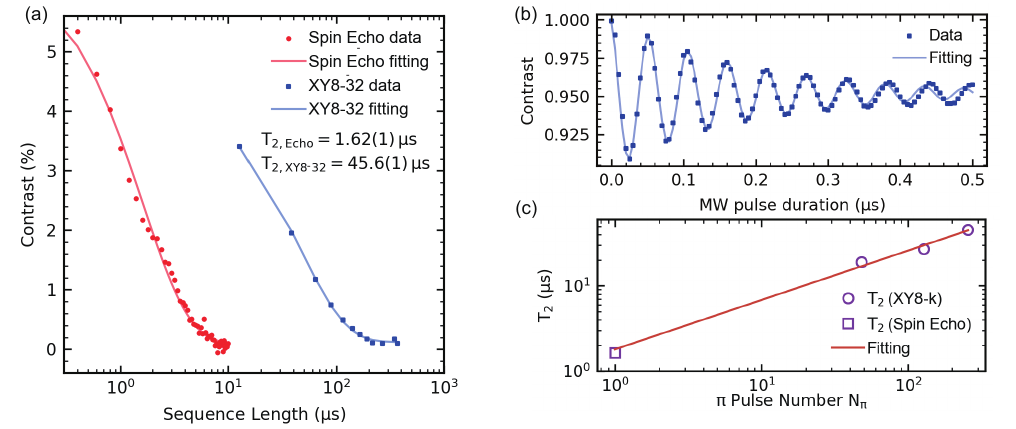}
    \caption{
    Supporting data and fitting results for ``Shallow NV'' parameters listed in Table II of the main text.  (a)  $T_2$ measured using Hahn echo ($T_{2\text{Echo}}=1.62(1)$)and XY8-32  ($T_{2\text{Saturated}}=45.6(1)$) sequences, with associated exponential fitting. (b) Rabi oscillation measurements exhibit $\approx$ 9\% contrast. (c) Experimentally determined echo and XY8-k $T_2$ with respect to $\pi$ pulse number $N_\pi$. The fitted slope is $s$=0.58(9). 
    }

    \label{shallowNV}
\end{figure*}

The diamond used to determine ``Shallow NV'' parameters is a CVD grown, isotopically purified (99.6\% $^{12}$C) [110] diamond from Element Six. NV centers are created with 6 keV $^{14}$N+ irradiation at a dose of 2$\times10^{13}\,$cm$^{-2}$ followed by annealing at 800$^\circ$C, producing a 2D NV density of about 3.5$\times$10$^{11}\,$cm$^{-2}$, which corresponds to a NV concentration of about 0.6 ppm. This irradiation process produces a $\sim$ 10$\,$nm thick NV layer centered at a depth greater than 10$\,$nm (see \cite{pham_nmr_2016} for a demonstration of how SRIM underestimates depth). $T_2$ measurements for this sample are performed using similiar techniques to those discussed above for the ``Bulk NV'' sample, with example results shown in Fig. \ref{shallowNV}.

\subsubsection{``Single NV'' Parameters}

Single NV parameters are based on typical values reported by QZabre Ltd \cite{qzabre_2025} for their single NV scanning tips (Table \ref{singleNVtable}). For the purpose of this paper, we select the value within each range that predicts the best sensitivity. The exponential scaling factor $s$ is assumed to be 0.5. 
\begin{table}
\begin{ruledtabular}
\begin{tabular}{|l|l|}
\hline
Contrast (at 10 MHz line width) & 15\% - 27\% \\
\hline 
Saturation count rate (0.75 NA) & 0.6 - 1.2 MCts/s \\
\hline 
Nominal NV depth & 10 nm \\
\hline 
T\textsubscript{2}/T\textsubscript{2}* (single Hahn echo/XY) & \specialcell{4 - 1 $\mu$s \\ (average values)} \\ 
\hline 
Nominal cw-ODMR sensitivity (Q2-Q8) & 1 - 7 $\mu$T/sqrt(Hz) \\ 
\hline 
Brightness (at max. contrast) & 100 - 1000 kcts/s\\ 

\end{tabular}
\end{ruledtabular}
\caption{\label{singleNVtable} Single NV parameters provided for QZabre scanning tips.}
\end{table}

\section{Sensitivity optimization}

For AC sensitivity calculations, the PL readout time $t_R$ is varied to optimize sensitivity $\eta$.  In Equation 1 of the main text, $t_R$ appears (explicitly or implicitly) in the following terms:
\begin{equation} 
\sqrt{1+\frac{1}{C^2 n_{avg}}}\sqrt{\frac{t_I+\tau+t_R}{\tau}}.
\end{equation}
We assume that the average AC measurement PL contrast follows an exponential decay model:
\begin{equation}
C = \frac{1}{t_R}\int_0^{t_R}e^{-t/t_I}dt = \frac{t_I}{t_R}(1-e^{-t_R/t_I}).
\label{contrast_Eq}
\end{equation}
The readout time $t_R$ is thus optimized by solving the following transcendental equation  numerically: 

\begin{multline}
    \hat{t}_R = \text{argmin}_{t_R}\sqrt{\frac{t_I^2 \left(e^{\frac{t_R}{t_I}} - 1\right)^2 + t_R \left(e^{\frac{2 t_R}{t_I}}\right)}{t_I^2 \left(e^{\frac{t_R}{t_I}} - 1\right)^2}}\\
 \cdot \sqrt{\frac{t_I + t_R + \tau}{\tau}}
\end{multline}

\begin{figure*} [h]
    \includegraphics[width=1\linewidth]{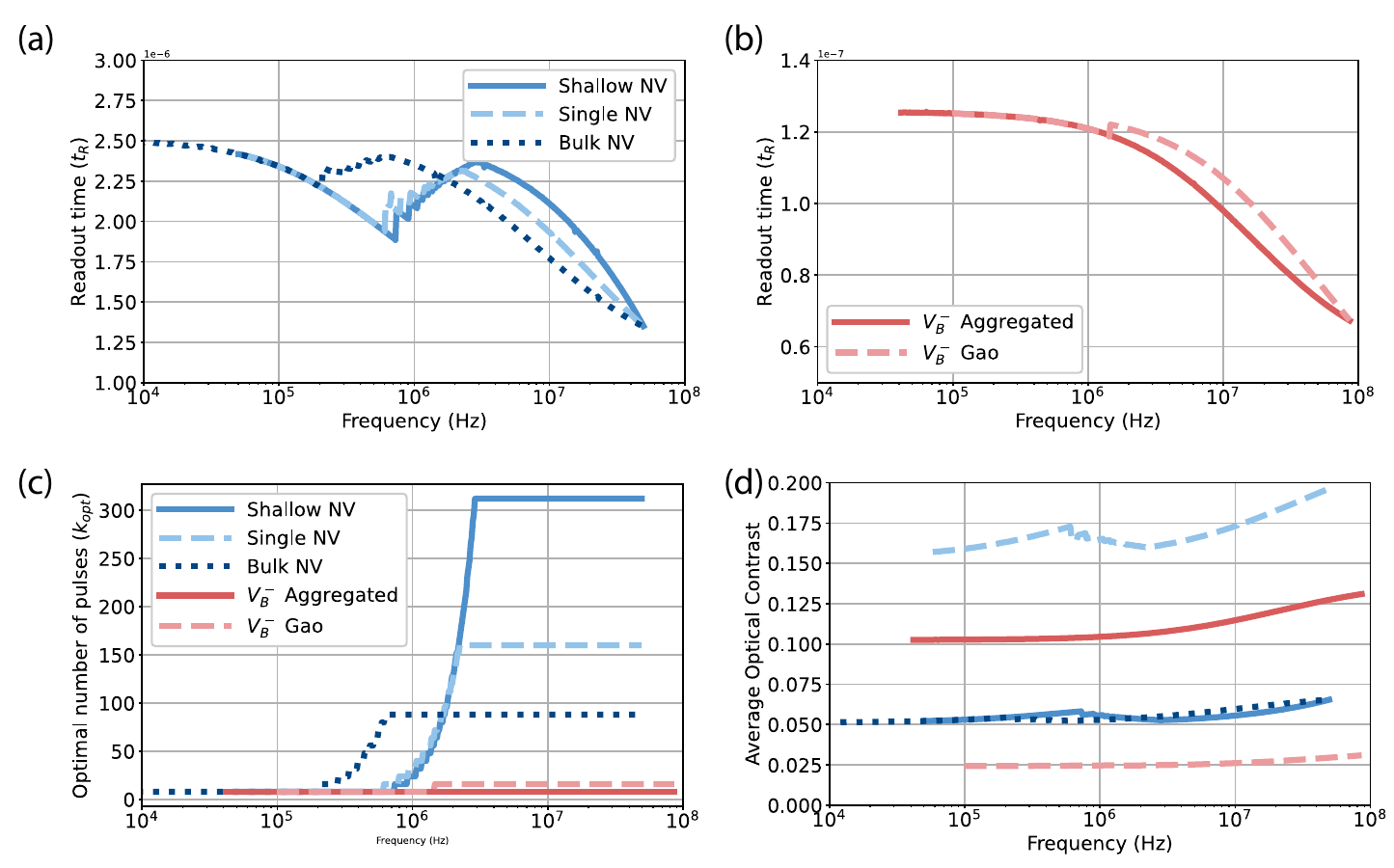}
    \caption{\label{sens_params}Dynamic parameter values vs.\ signal frequency used in AC sensitivity calculations (Fig. 3 in the main text). Only points associated with volume-normalized sensitivity $\eta<1\,$mT Hz$^{-\frac{1}{2}}\mu\text{m}^{\frac{3}{2}}$ are shown. (a) Optimal readout time, $t_R$ vs.\ signal frequency for NV systems. Step features arise from restricting the number of pulses to multiples of 8 for XY8-k sequences. (b) Optimal readout time, $t_R$ vs.\ signal frequency for $V_B^-$ systems. As the saturated $T_2$ is closer to the Hahn echo $T_2$, there is less frequency dependence than for NV systems. (c) Optimal number of pulses, $k_{opt}$ vs.\ signal frequency. The visible step features are due to enforced rounding to multiples of 8 for XY8-k sequences. The eventual plateau represents the point at which maximum $T_2$ is saturated, which differs for each system. (d) Average optical AC measurement PL contrast over the course of the readout period vs signal frequency (legend is shared with c). This parameter depends on the base AC contrast, as well as  $T_2$, $t_R$, and the re-initialization time, $t_I$ (see Eq. \ref{contrast_Eq}) }
\end{figure*}

\begin{figure} [h]
    \includegraphics[width=1\linewidth]{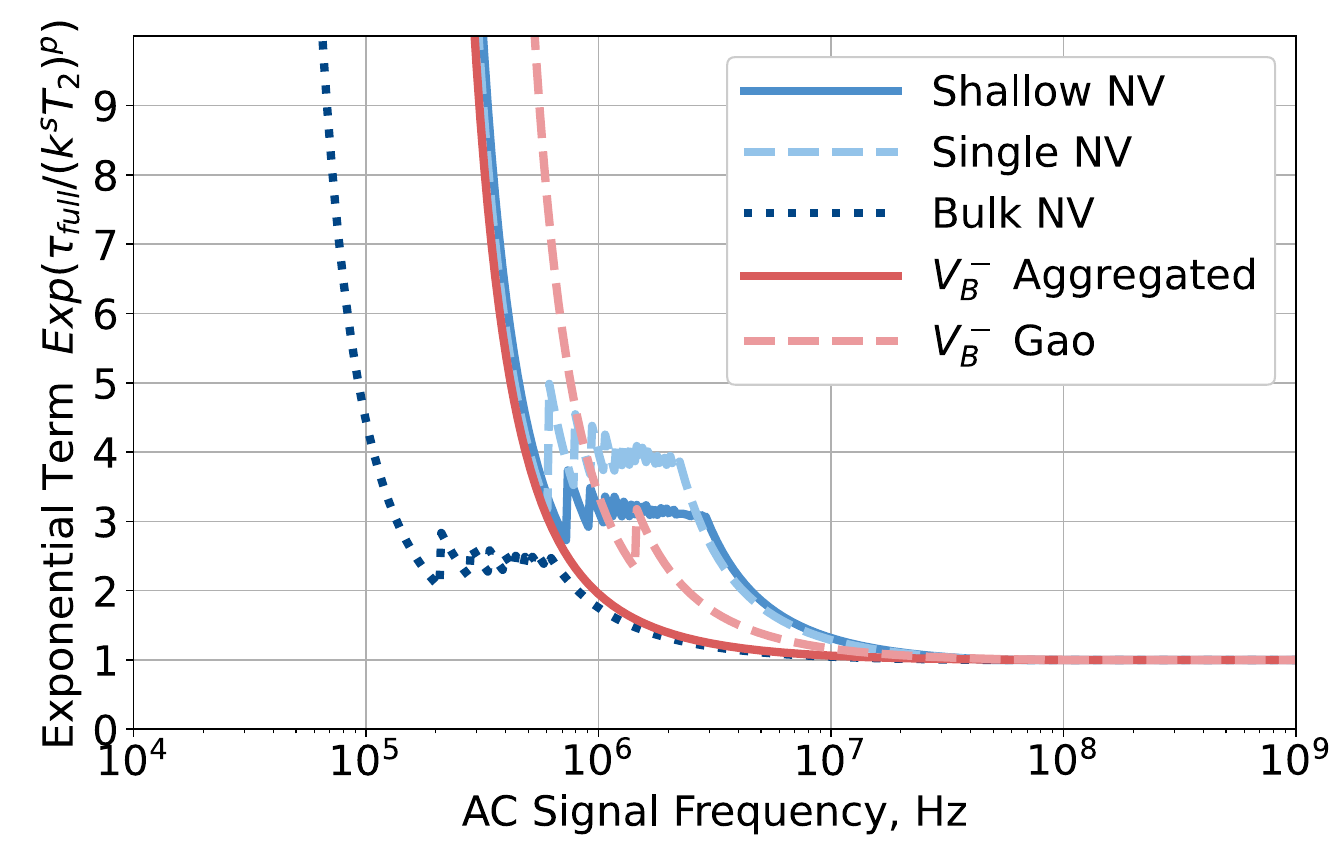}
    \caption{\label{exp_scale}Scaling with AC signal frequency of the exponential term of Equation 1 of the main text.}
\end{figure}

For the different sensor modalities considered in this work, dynamic values for $t_R$, $C$, and $k_{opt}$ (see Equation 2 of the main text) are shown vs. AC signal frequency in Fig. \ref{sens_params}; and scaling with AC signal frequency of the exponential term in Equation 1 of the main text is shown in Fig. \ref{exp_scale}.

\section{Geometry Factor and Sensitivity Scaling}

In this section, we follow the argument in \cite{bruckmaier_geometry_2021} and highlight how sample geometry has a large impact on the strength of an NMR signal detected by optically-active spin defects in a solid. 

Consider a typical, micron-scale NMR experiment where the bias magnetic field unit vector, $\hat{B}_0$, is aligned with the principal axis of the defect. The strength of the NMR signal detected by a defect is determined by the dipolar field profile of the sample spins, located above the surface of the solid-state sensor. The signal is detected via modulation of the defect's resonance frequency and its amplitude is therefore given by the projection of the dipolar field along $\hat{B}_0$. For a spin $j$, the 
\begin{equation}
B_{Signal,j} = \vec{B}_{j}\left(r\right)\cdot \hat{B}_0.
\label{eqs1}
\end{equation}
where $\vec{B}_{j}(r)$ is the dipolar magnetic field produced by (thermally-polarized or hyperpolarized) sample spin $j$.  Assuming the sample is homogeneous, and converting from discrete sample spins to a spin density, the volume average of \ref{eqs1} is:
\begin{equation}
B_{Signal} = \rho \frac{\mu_0 |\vec{m}|}{4 \pi} \int \frac{3(\hat{r}\cdot\hat{m})(\hat{r}\cdot\hat{B_0})-\hat{m}\cdot\hat{B_0}}{r^3}dV.
\end{equation}
Here, $\rho$ is the volumetric density of polarized sample nuclear spins and $\vec{m}$ is the magnetization vector of the sample spins.  
The geometry factor $G$ is defined as the dimensionless integral:
\begin{equation}
G = \int \frac{3(\hat{r}\cdot\hat{m})(\hat{r}\cdot\hat{B_0})-\hat{m}\cdot\hat{B_0}}{r^3}dV.
\end{equation}
For an NMR experiment detecting the transverse component of the sample magnetization (such as CASR), $\hat{B}_0$ and $\hat{m}$ are orthogonal, allowing for the geometry factor to be evaluated for a hemispherical sample volume above the surface by defining $\hat{B}_0 = \sin \alpha\hat{x} + \cos \alpha \hat{z}$ and $\hat{m} = -\cos \alpha \hat{x} + \sin \alpha \hat{z}$. Here $\alpha$ is the defect orientation angle away from $\hat{z}$.  Evaluation of S3 then results in \cite{bruckmaier_geometry_2021}:
\begin{equation}
G_{transverse} = \pi \sin(2\alpha)[1 + \frac{1}{2}\left(\epsilon\right)^{3}-\frac{3}{2}\left(\epsilon\right)],
\end{equation}
where $\epsilon =d_{r}/R_{max}$ is the ratio of the largest sample volume radius to the depth of the defect sensor. For most experiments $\epsilon$ is small and these terms can be ignored. For longitudinal NMR signals such as those in an AERIS-type experiment, $\hat{B}_0$ and $\hat{m}$ are aligned, resulting in the following geometry factor in the small $\eta$ limit:
\begin{equation}
G_{longitudinal} = \pi \left(\cos (2\alpha)+\frac{1}{3}\right).
\end{equation}
NV diamonds cut in the typical [100] or [110] orientation have tetrahedral symmetry such that $\alpha=54.7^\circ$, the magic angle, and $G_{longitudinal}\approx 0$. $V_B^-$ defects, however, operate with $\alpha = 0$ or $\alpha= \pi$, corresponding to local maxima of $G_{longitudinal}$. This result indicates that hBN defects are optimally sensitive to longitudinal NMR signals. While other diamond cuts, such as [111], also have $\alpha = 0$, shallow NVs in these diamonds have not yet been reported.

\begin{figure}[h]
    
    \includegraphics[width=1\linewidth]{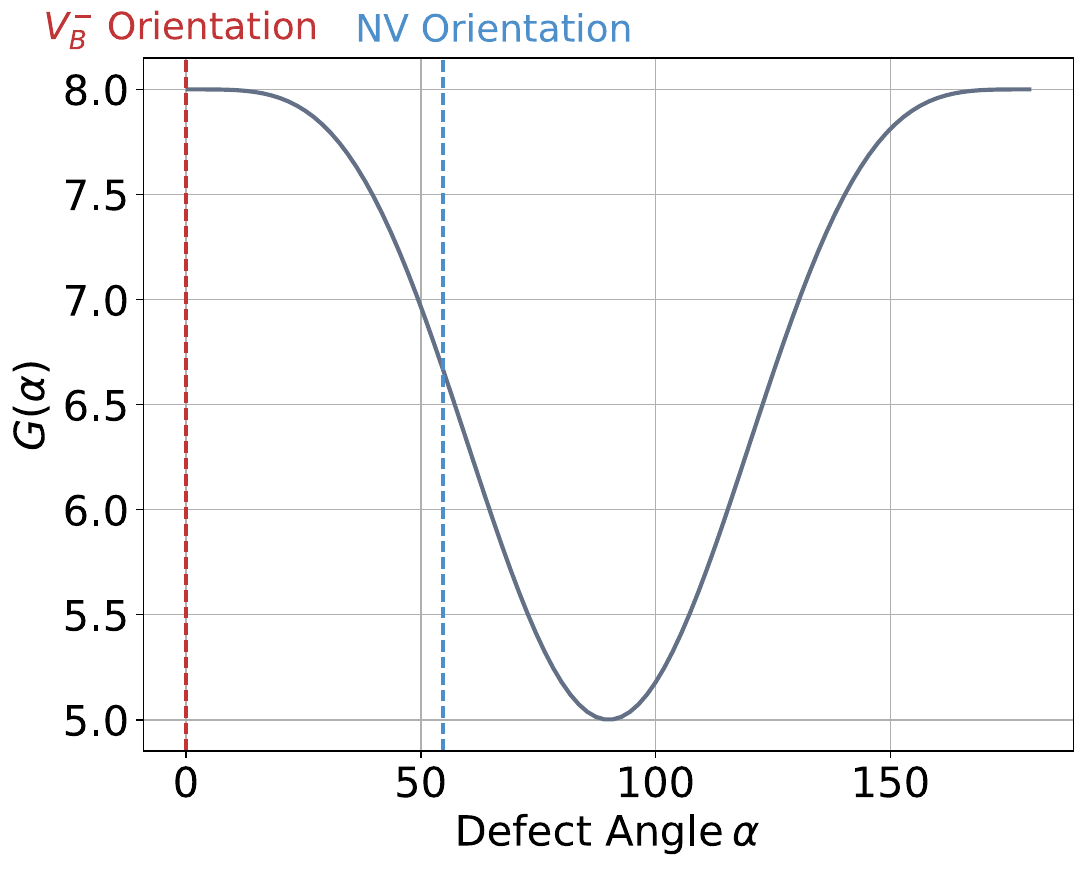}
    \caption{\label{statgeo} Geometry factor vs.\ defect angle for statistical polarization NMR measurements. $V^{-}_{B}$ defects are oriented in a more favorable direction than [110] or [100] cut NV centers.  $G(\alpha)$ is directly calculated from the analytic model discussed in \cite{pham_nmr_2016} to be $G(\alpha) = 8-3\sin^4(\alpha)$. }

\end{figure}

For the nanoscale regime where statistical spin polarization dominates, we use a geometry factor calculation from \cite{pham_nmr_2016}, again finding a substantial advantage for $V_B^-$ defects relative to NVs. See Fig. \ref{statgeo}. 

\section{Calculating SNR for a Given Sensor AC sensitivity}

A general expression for the root-mean-square AC magnetic field amplitude from a statistically polarized sample at the nanoscale, as developed in \cite{pham_nmr_2016}, is as follows:
\begin{equation}
B^2_{rms} = \rho \left(\frac{\mu_0 \hbar \gamma_n}{4 \pi}\right)^2 \left(\frac{\pi G(\alpha)}{128 d_{r}^3}\right)
\end{equation}
Setting $B_{rms} = \text{SNR}\times\eta$, where $\eta$ is the defect AC sensitivity, and solving for \text{SNR}:
\begin{equation} \label{supplementalSNREquation}
\text{SNR} = \frac{(\mu_0 \hbar \gamma_n)}{32 \eta}\sqrt{\frac{G(\alpha)\rho}{2\pi d^3_r}}
.\end{equation}
For thin layer samples, $1/d_r^3$ becomes $\left( \frac{1}{(d_r)^3}-\frac{1}{(d_r + h)^3}\right)$ where $h$ is the layer thickness.  Equation \ref{supplementalSNREquation} can thus be adjusted to accommodate this factor:
\begin{equation}
\text{SNR}_{\text{flake}} = \frac{(\mu_0 \hbar \gamma_n)}{32 \eta}\sqrt{\frac{G(\alpha)\rho}{2\pi}\left( \frac{1}{(d_r)^3}-\frac{1}{(d_r + h)^3}\right)}
.\end{equation}
For bulk samples, the active NV layer is too thick for SNR to be well-represented at one depth.  To compensate, we express the SNR for the bulk sample as an average value (since noise is approximately constant for all defects in the integration);

\begin{multline}
\text{SNR}_{\text{bulk}} = \frac{1}{d_{\text{max}}-d_{\text{min}}}\int_{d_{\text{min}}}^{d_{\text{max}}} \text{SNR}(d_r) \text{d}d_r = \\  -\frac{\mu _0 \hbar  \gamma _n \left(\sqrt{\frac{1}{d_{\max }^3}}
   d_{\max }-\sqrt{\frac{1}{d_{\min }^3}} d_{\min }\right) \sqrt{\rho
    G(\alpha )}}{16 \sqrt{2 \pi } \eta  \left(d_{\max }-d_{\min
   }\right)}  
\end{multline}

Performing the same integration for the flake is fairly demanding, requiring invocation of elliptic functions or complex contours.  To simplify the system we expand around small flake thickness \textit{h} to third order, resulting in the following expression:
\begin{multline}
\text{SNR}_{\text{flake, bulk}} 
 = \frac{\mu _0 \hbar  \gamma _n \sqrt{\rho  G(\alpha )}}{192 \sqrt{6 \pi } \eta  \left(d_{\max }-d_{\min }\right)}\\
   (\frac{\sqrt{\frac{h}{d_{\max }^4}} \left(9 h d_{\max }-18
   d_{\max }^2-7 h^2\right)}{d_{\max }}\\+\frac{\sqrt{\frac{h}{d_{\min
   }^4}} \left(-9 h d_{\min }+18 d_{\min }^2+7 h^2\right)}{d_{\min
   }})
\end{multline}

\section{Additional Back-Action Simulations}

\begin{figure*} [h]
    \includegraphics[width=1\linewidth]{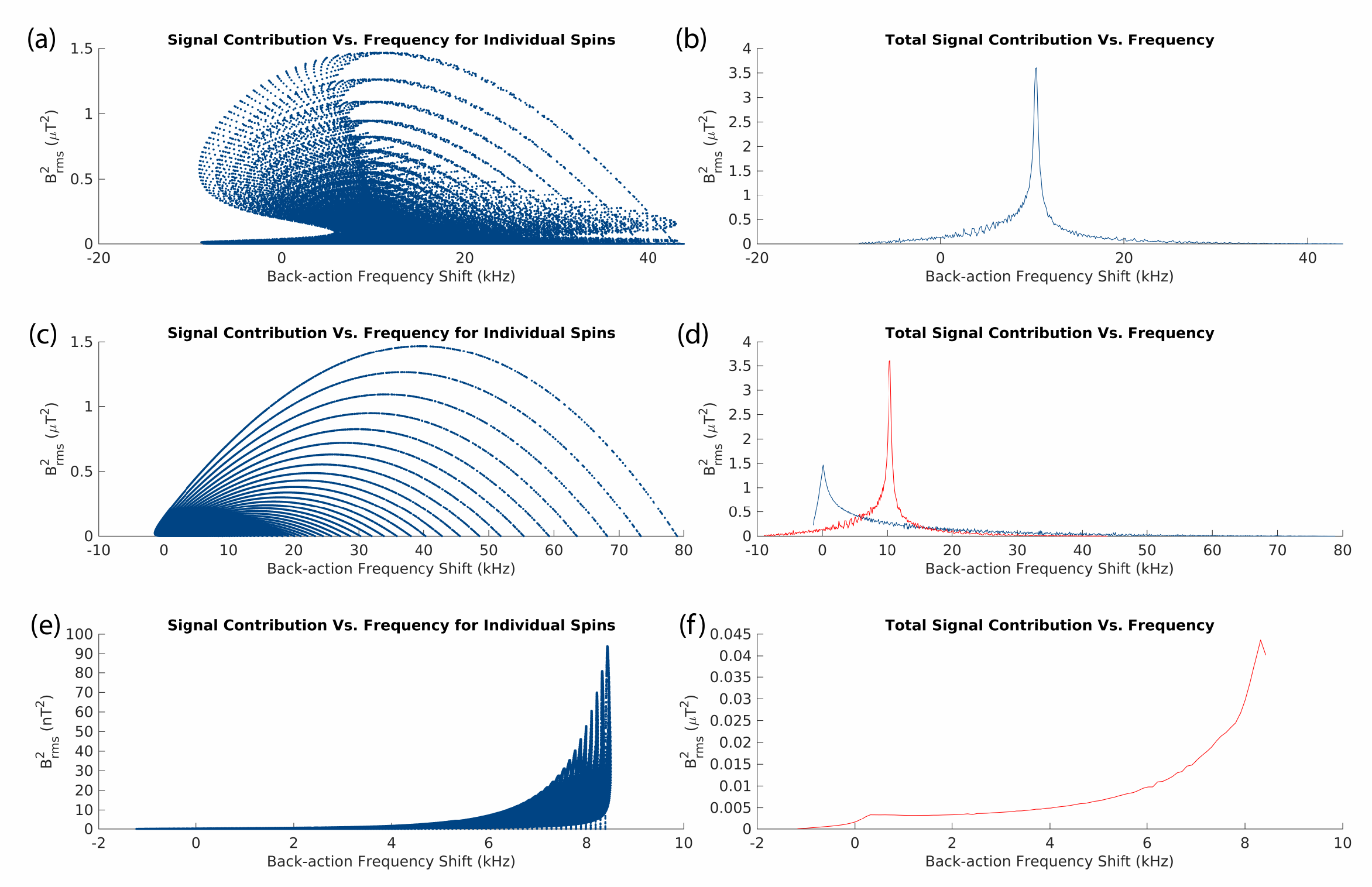}
    \caption{\label{backactionSI} Model calculations indicate that dense $V_B^-$ ensembles (236 ppm) result in larger back-action-induced NMR lineshape broadening than for a single $V_B^-$ defect for all but the shallowest defect depths. In the model calculations, a sample with proton density $\rho = 64 \text{ protons/nm}^3$ is confined to a $(4d)^3$ volume above the surface, where $d$ is the sensor depth. (a) Back-action shift of each sample spin NMR signal caused by close proximity to a $V_B^-$ electronic spin ensemble (average $d=1$\,nm). (b) Total NMR signal measured by the same $V_B^-$ system, resulting in $\sim$800 Hz broadening and $\sim$10 kHz frequency shift. (c) Back-action shift of each sample spin signal caused by close proximity to a single, $d=1$\,nm deep $V_B^-$ electronic spin. (d) Comparison of broadening effects explored above, as measured by a $V_B^-$ ensemble (red) and single $V_B^-$ center (blue)  (e) Back-action shift of each sample spin signal caused by proximity to a $V_B^-$ electronic spin ensemble (average $d=5$\,nm). (f) Total NMR signal measured by the same $V_B^-$ system, resulting in $\sim$1 kHz broadening and $\sim$8 kHz frequency shift. 
    }
\end{figure*}

To supplement Figs. 5 and 6 from the main text, similar back-action simulations are performed for dense $V_B^-$ ensembles (236 ppm, giving 1.4 nm lateral spacing for 10 layers of hBN).  NMR lineshape broadening from back-action is found to be less dependent on defect depth for $V_B^-$ ensembles than for single defect systems, but the shape of the spectral broadening varies greatly (Fig. \ref{backactionSI}).

\begin{figure*} [h]
    \includegraphics[width=1\linewidth]{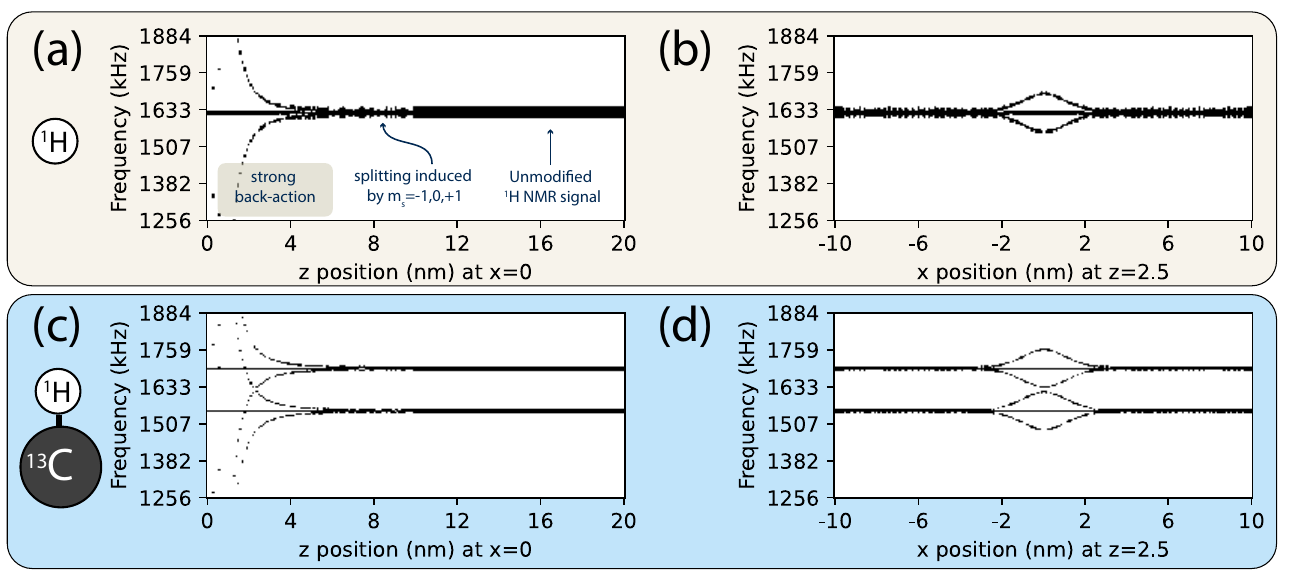}
    \caption{\label{backaction}Simulated CASR NMR frequencies at the nanoscale for a single $V_B^-$ sensor and one or two nuclear spins at defined locations.  Back-action between the sensor electronic and sample nuclear spins induces frequency shifts in NMR spectral peaks as a function of relative sensor/sample separation and the $V_B^-$ spin state ($m_s$=-1,0,+1), becoming strong at few nanometer distances.  In all results shown here, the $V_B^-$ sensor is located at z = –2.5 nm (below the hBN surface) and at x = 0; and the bias magnetic field = 0.1\,T is aligned with the $V_B^-$ quantization axis along z (perpendicular to the hBN surface). (a) One proton spin at x = 0, as a function of distance z perpendicular to the hBN surface.  (b) One proton spin at z = 2.5 nm, as a function of distance x parallel to the hBN surface.  (c) One proton spin coupled to one 13C nuclear spin at x = 0, as a function of distance z perpendicular to the hBN surface.  (d) One proton spin coupled to one 13C nuclear spin at z = 2.5 nm, as a function of distance x parallel to the hBN surface. }
\end{figure*}

\begin{figure} [h]
    \includegraphics[width=1\linewidth]{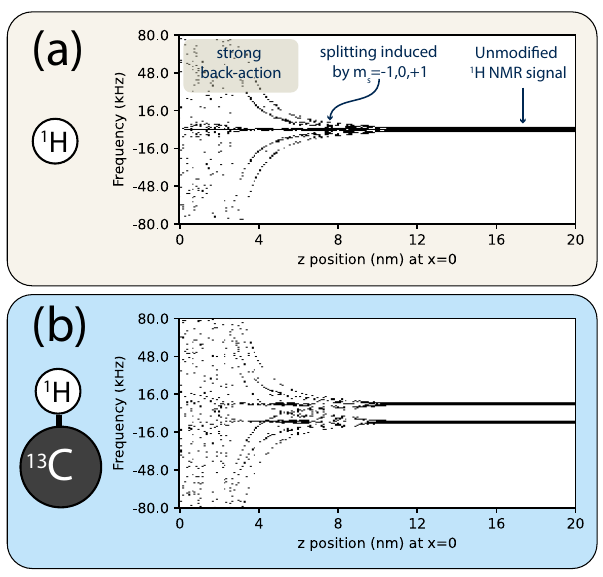}
    \caption{\label{backaction_ensemble}Simulated NMR frequencies at the nanoscale for the summed interaction between four $V_B^-$ electronic spin sensors and one or two nuclear spins at defined locations.  Back-action between the sensor electronic and sample nuclear spins induces frequency shifts in NMR spectral peaks as a function of relative sensor/sample separation.  In all results shown here, the nuclear spin sample is located at the origin, (x,y,z) = (0,0,0); the four $V_B^-$ sensors are randomly located at or below the hBN surface, consistent with a mean spacing of 5 nm, and their x or z positions are then varied; and the bias magnetic field = 0.1\,T is aligned with the $V_B^-$ quantization axis along z (perpendicular to the hBN surface).  At very short standoff distances, heavy aliasing of simulated $V_B^-$ NMR measurements occurs due to strong back-action effects.  (a) One proton spin as a function of relative distance z of the four $V_B^-$ sensors perpendicular to the hBN surface. (b) One proton spin coupled to one $^{13}$C nuclear spin (2.25 ppm scalar coupling) as a function of relative distance z of the four $V_B^-$ sensors perpendicular to the hBN surface.}
\end{figure}

We also assess the effect of quantum mechanical back-action modifying the $V_B^-$ sensor and sample spin resonance properties through simulation.  Using the NMR simulation tool Spinach \cite{hogben_spinach_2011}, we model two systems: a single $^1$H (proton) nuclear spin measured at 200 separate, fixed locations varying in $x$ or $z$ relative to the $V_B^-$ sensor (with 1 Angstrom resolution), as well as a formic acid molecule ($^1$H$^{13}$C OO$^1$H ) with scalar coupling between $^1$H and $^{13}$C measured at the same set of locations. This comparison allows us to examine the impact of back-action effects on signals measured with $V_B^-$ NMR. We find that the distance from the defect can strongly affect the NMR spectra, dependent on the $V_B^-$ electronic spin states $m_s = -1, 0, +1$; as seen in Fig. \ref{backaction}. Within 10\,nm, back-action introduces additional splittings in the NMR spectra, whereas beyond $\sim$10 nm, unperturbed (no back-action) NMR spectra are found.

While the above analysis only considers the local environment of one nuclear spin $\hat{I}$, we can use it to estimate the behavior of a multi-spin sample detected by a small $V_B^-$ ensemble.  For each near-surface nuclear spin $\hat{I}_i$, the $V_B^-$ ensemble will detect NMR spectra with significant back-action-induced features.  However, due to the nuclear spins being statistically polarized and thus having random phases, the back-action-induced NMR spectral features from multiple nearby nuclear spins are expected to average out, to leading order, producing a very broad net NMR spectrum ($\sim$1 kHz).  The net NMR signal from sample nuclear spins farther than 5-10$\,$nm from the $V_B^-$ sensor spins is not estimated to average out, given the minimal back-action effects at these distances.  Combining all discussed effects, we expect to observe NMR spectral peaks with broad bases and narrow tips in measurements involving multiple sample spins; the broad base would reflect the mixture of back-action distances involved, and the narrow tips would comprise both the $m_s=0$ signal and the low-interaction, aggregate sample signal.

Fig. \ref{backaction}a is produced by initializing an electronic spin ('E3' in Spinach) and a nuclear spin at the respective coordinates: (0,0,0), (0, 0, \textit{z}) where \textit{z} is varied between 0 and 20\,nm.  The $V_B^-$ electronic g-factor $=$ 2.001 \cite{gracheva_symmetry_2023}.  The $^1$H chemical shift $=$ 3.25 ppm.  Both spins are initialized by $\pi/2$ Bloch rotations prior to the NMR measurement, which is simulated for 0.2 seconds with 5000 time steps.  Fig. \ref{backaction}c adds a $^{13}$C spin at coordinate (0, 1.1, \textit{z}) with chemical shift of 2.25 ppm. Figs.  \ref{backaction}b and  \ref{backaction}d show similar results for varying the $x$ position of the target nuclear spin or spins, with $z$ fixed at 2.5$\,$nm. 

Simulations are also performed with the hBN nearest-neighbor nuclear spin environment around the $V_B^-$ sensor, including hyperfine coupling and robust relaxation operators.  These simulations include 1 $V_B^-$ electronic spin, 3 $^{14}$N nuclear spins, and 1$^1$H proton spin, with the $^{14}$N chemical shift $=$ 61.2 ppm.  Supplied system coordinates are (0,0,0), (1.25, 0.72, 0), (-1.25, 0.72,0), (0, -1.45, 0), and (0, 0, \textit{z}) in Angstroms, where $z$ is varied between 0 and 20 nm.  

For the simulations summarized in Figs. \ref{backaction_ensemble}a-b, four $V_B^-$ electronic spins are initialized at locations (0, 0, $z$), (-16.6, -43.7, 8.4), (30.0, 52.8, -13.3), and (-4.3, -0.5, 5.8) in Angstroms; and their $x$ or $z$ positions are uniformly varied relative to the nuclear spin target. Nuclear spin properties and respective coordinates are inherited from values used for Fig. \ref{backaction}a-d.

In all simulations, the interaction couplings are calculated dynamically by Spinach based on specified parameters. Calculations are performed using a Nvidia RTX A6000 GPU with ``sys.enable={`greedy',`gpu'}.
\clearpage
\bibliography{main}

\end{document}